\documentclass[usenatbib]{mn2e}

\usepackage{graphicx}
\usepackage{txfonts}
\usepackage{url}
\usepackage{natbib}

\title[The IAC Stripe 82 Legacy Project]{The IAC Stripe 82 Legacy Project: a wide-area survey for faint surface brightness astronomy}

\author[J\"urgen Fliri and Ignacio Trujillo]{J\"urgen Fliri$^{1,2}$ \thanks{E-mail:
jfliri@iac.es; trujillo@iac.es} and Ignacio Trujillo$^{1,2}$\\
$^{1}$Instituto de Astrof\'{\i}sica de Canarias, c/ V\'{\i}a L\'actea s/n, 
E-38205, La Laguna, Tenerife, Spain\\
$^{2}$Departamento de Astrof\'{\i}sica, Universidad de La Laguna, E-38206, La 
Laguna, Tenerife, Spain}
\begin{document}

\date{}

\pagerange{\pageref{firstpage}--\pageref{lastpage}} \pubyear{2013}

\maketitle

\label{firstpage}

\begin{abstract}

We present new deep co-adds of data taken within Stripe 82 of the Sloan
Digital Sky Survey (SDSS), especially stacked to reach the faintest
surface brightness limits of this data set. Stripe 82 covers 275
$\deg^2$ within --50\degr $\leqslant$ RA $\leqslant$ +60\degr and
--1.\degr25 $\leqslant$ Dec. $\leqslant$ +1.\degr25.  We discuss
the steps of our reduction which puts special emphasis on preserving
the characteristics of the background (sky + diffuse light) in the
input images using a non-aggressive sky subtraction strategy. Our
reduction reaches a limit of $\sim$28.5 mag arcsec$^{-2}$ (3$\sigma$,
10$\times$10~arcsec$^2$) in the $r$ band. The effective surface
brightness limit (50\% completeness for exponential light
distribution) lies at $<\mu_e(r)>$$\sim$25.5~mag arcsec$^{-2}$. For
point sources, we reach 50\% completeness limits (3$\sigma$ level) of
(24.2, 25.2, 24.7, 24.3, 23.0)~mag in $(u,g,r,i,z)$. This is between 1.7
and 2.0 mag deeper than the single-epoch SDSS releases.  The
co-adds show point spread functions (PSFs) with median full width at
half-maximum values ranging from 1 arcsec in $i$ and $z$ to 1.3 arcsec
in the $u$ band.  The imaging data are made publicly available at
http://www.iac.es/proyecto/stripe82.  The release includes deep co-adds
and representations of the PSF for each field. Additionally, we
provide object catalogues with stars and galaxies confidently
separated until $g$$\sim$23~mag. The IAC Stripe 82 co-adds offer a
rather unique possibility to study the low surface brightness
Universe, exemplified by the discovery of stellar streams around
NGC0426 and NGC0936.  We also discuss further science cases like
stellar haloes and disc truncations, low surface brightness galaxies,
the intra-cluster light in galaxy clusters and the diffuse emission of
Galactic dust known as Galactic Cirrus.
\end{abstract}



\section{Introduction}

There is a huge amount of astrophysical phenomena that remain still
barely studied due to the lack of large (several hundreds of square
degrees), multiwavelength and deep ($\mu_V$$>$28 mag~arcsec$^{-2}$)
optical surveys. These unexplored astrophysical events are those which
are very subtle and extend over large areas of the sky. For instance,
little is known about the connection of the so-called `optical cirrus'
or diffuse light of our Galaxy \citep{deVries1985,Paley1991}
and the dust filamentary structure observed in the far-infrared
full-sky surveys \citep[i.e.][]{Low1984}. Also, only a relative small
number of nearby galaxies have been probed with enough depth
\citep[e.g.][]{Ferguson2002,Mouhcine2005,McConnachie2006,Mouhcine2007,Martinez2008,Radburn-Smith2011,Bakos2012,Trujillo2015}
to explore the cosmological predictions
\citep[e.g.][]{Bullock2005,Cooper2010,Font2011,Tissera2013} about the
formation of the faint stellar haloes, tidal streams and ultra-faint
satellites surrounding these objects.  Similarly, only a handful of
nearby galaxy clusters
\citep[e.g.][]{Rudick2010,Giallongo2014,Montes2014} have been observed
with enough depth to start understanding the intracluster light (ICL)
expected from a hierarchical assembly \citep[e.g.][]{Contini2014} of
these cosmic structures.

In the last few years there have been a few dedicated surveys designed
to explore some of the above astronomical questions. To name a few:
(1) a survey to explore stellar tidal streams in spiral galaxies
of the Local Volume \citep{Martinez2010}. This survey uses small
(d=0.1-0.5m) robotic telescopes to observe stellar streams down to
$\mu_V$$\sim$28.5~mag arcsec$^{-2}$, establishing a first
classification scheme for the tidal features. (2) The Pan-Andromeda
Archaeological Survey \citep[PAndAS;][]{McConnachie2009} conducted at
the 3.6m Canada--France--Hawaii Telescope whose main aim is the
exploration of the stellar halo surrounding our neighbouring Andromeda
galaxy.  This survey explores 400 square degrees reaching the
following optical depth for point sources: $g$=25.5 and $i$=24.5 mag
(S/N=10).  (3) Using the same telescope, another deep and large (104
square degrees) survey is the Next Generation Virgo Cluster Survey
\citep[NGVS;][]{Ferrarese2012}. Among its observational goals is the
exploration of the ICL in our closest galaxy cluster.  The depth
reached with this survey is the following: $u$=26.3 (S/N=5), $g$=25.9
(S/N=10), $r$=25.3 (S/N=10), $i$=25.1 (S/N=10), $z$=24.8 mag (S/N=5)
[point sources].  (4) The Dragonfly telescope has joined recently in
the initiative of exploring large areas of the sky to faint surface
brightness limits.  This telescope uses an array of telephoto cameras
reaching a surface brightness limit of $\mu_g$$\sim$29.5~mag
arcsec$^{-2}$ on scales of 10$\times$10 arcsec$^{2}$
\citep{Merritt2014}. However, there has not been yet a plan for a
general multipurpose deep survey expanding over a large area of the
sky that allows a systematic analysis of the phenomena stated in the
previous paragraph\footnote{A survey that potentially could have been
  used for the above purposes would be the Canada-France-Hawaii Legacy
  Survey \citep[CFHTLS;][]{Cuillandre2012}, in particular with its
  Wide Synoptic branch (155 square degrees with the following depth:
  $u$=25.3, $g$=25.5, $r$=24.8, $i$=24.4, $z$=23.5 mag; 50\%
  completeness for point sources).  Unfortunately, the pipeline used
  for the reduction of this data set removes the low surface brightness
  features, producing obvious `holes' around the brightest extended
  galaxies in the images.}. Nonetheless, many of the above issues
could be addressed with a proper reduction of the data collected along
the celestial equator in the Southern Galactic Cap of the Sloan
Digital Sky Survey \citep[SDSS;][]{York2000}, known popularly as the
`Stripe 82' survey \citep{Jiang2008,Abazajian2009}.

The Stripe 82 survey is a 2.\degr5 wide region along the celestial
equator (--50\degr $\leqslant$ RA $\leqslant$ +60\degr, --1.\degr25
$\leqslant$ Dec.  $\leqslant$ +1.\degr25 for a total of 275 square
degrees). This region of the sky has been imaged repeatedly
approximately 80 times in all the five SDSS filters: $ugriz$. The
Stripe 82 area is a perfect piece of the sky for exploring many of the
astrophysical phenomena described above. First, is accessible for the
vast majority of ground-based facilities, helping to create ancillary
spectroscopic and photometric observations if needed. Secondly, for
analysing the optical emission of the dust of our own Galaxy, it
covers regions from low to high Galactic extinction \citep{Schlegel1998}.

Around a third of all the available SDSS data in the Stripe 82 area
(123 of 303 runs\footnote{An SDSS run is a single continuous drift scan
  obtained on a single night.}) were originally combined by
\cite{Annis2014}. The goal of that co-addition was `to use this deep
survey to understand the single pass data at its limits and to do
science at fainter magnitudes or correspondingly higher redshifts'
\citep{Annis2014}.  This stack produced the following depths (50\%
completeness limit for point sources): $u$=23.6, $g$=24.6, $r$=24.2,
$i$=23.7, $z$=22.3 mag with a median seeing of $\sim$1.1 arcsec.
Depending on the band, this is around 1--2 mag deeper than SDSS
single-pass data. In a second reduction of the Stripe 82 data,
\citet{Jiang2014} combined all available runs; the major differences
to \cite{Annis2014} were the amount of data used and a different
treatment of the sky background using a 2D background
modelling. Depending on the band, \citet{Jiang2014} used between 75
and 90 per cent of the data for their co-adds, reporting a gain of 0.3--0.5 mag
in depth compared to the \cite{Annis2014} reduction.

The science foreseen for the previous co-additions of the Stripe 82
survey has been focused on `point sources science' (i.e. Galactic
structure, photometric redshift computation, cluster finding, etc).
However, the real treasure of the Stripe 82 survey is not only related
to the detection of `point-like' sources but in a different unique
aspect: the exquisite surface brightness depth that a proper
combination of this sub-project of the SDSS can reach over a wide
area. In fact, one of the particular aspects of the SDSS survey is
that it has been conducted using drift-scan mode. This observational
technique has proved to be very efficient avoiding many of the
artefacts affecting the quality of the imaging.  Consequently, the
single-pass SDSS imaging is superb for studies of low surface
brightness features \citep[reaching down to $\sim$26.5 mag
  arcsec$^{-2}$ through direct detection (3$\sigma$, 10$\times$10
  arcsec$^2$) and $\sim$28 mag arcsec$^{-2}$ through profile averaging
  techniques;][]{Pohlen2006}. These are remarkable numbers taking into
account that the SDSS survey has been obtained with exposure
integrations of 53.9s using a 2.5m telescope. Considering that, on
average, any region of the sky in the Stripe 82 area has been observed
80 times, a simple calculation shows that an optimal combination
(i.e. assuming all single passes were of the same quality) of the
Stripe 82 would be able to reach $\sim$2.4 mag deeper than the regular
SDSS. In practice, as we will show, only around 2/3 of the full
available Stripe 82 data are useful (i.e.  with reasonable seeing and
darkness) for our purposes.

The goal of this paper is to provide to the entire astronomical
community with a new reduction of the Stripe 82 data, especially
designed to prevent the destruction of the low surface brightness
features in the co-addition of the data set. Moreover, our new
co-addition takes advantage of the full useful data available in the
Stripe 82 area, closely doubling the one third used by
\cite{Annis2014}. This allows us to go around 0.1--0.3 mag deeper than
before. Our data release includes not only $ugriz$ images and catalogues
of the Stripe 82 but also some ancillary products as stacked
point-spread function (PSF) stamps along the survey area plus an
`ultra-deep $r$-band' image of the survey obtained by the stacking of
the $g$, $r$ and $i$ co-added images. All our data are fully available
for the astronomical community through a dedicated
webpage at http://www.iac.es/proyecto/stripe82. Through the
paper, we will show some examples of the quality of the new reduction
and some of the science that can be conducted with this co-addition.
The paper is structured as follows: in Sec.~2 we introduce the data
used in our reduction of Stripe 82 while the reduction pipeline is
presented in Sec.~3. We assess the quality of the co-added data in
Sec.~4 and describe the contents of the data release in Sec.~5.
Section 6 offers some science cases which can be explored with the
Stripe 82 data. We end the paper with the conclusions in Sec.~7.

\section{Data}

Reduced images for all 303 runs covering Stripe82 are available as
part of the SDSS DR7 \citep{Abazajian2009} in a data base called Stripe
82. We have accessed these data through the Data Archive Server
(DAS). The quality of the 303 runs changes from optimal seeing, sky
brightness and photometric conditions for 84 runs obtained between
1998 and 2004 to 219 additional runs where the conditions were poorer
including worst seeing, bright moonlight and/or non-photometric
skies. These last runs were obtained from 2005 to 2007 as part of the
SDSS-II Supernovae Survey \citep{Frieman2008}.

SDSS data are taken in scans along lines of constant survey latitude
defining an SDSS strip. Two strips called strip `north' and `south' are
needed to fill the 2.\degr5 width of an SDSS stripe. The run coverage
of Stripe 82 is inhomogeneous over the survey area, but generally
rises from west to east. For the observations in the Stripe 82
southern strip, the difference in coverage between the western and the
eastern limit of the survey is about 25 per cent. The difference is smaller
for the northern strip with about 15 per cent. The co-addition that we present
in this paper uses all the data (i.e. the 303 available runs)
explained above after removal of some very bad data as we will explain
later on in Sec. \ref{coaddition}. This is a major difference compared to
\cite{Annis2014}. In fact, they combined only the data taken until
2005 (mostly because that data were not available when they were doing
their co-addition).  \cite{Annis2014} co-addition includes a total of
123 runs with every piece of the Stripe 82 area observed between 15 and
34 times. For comparison, any region of the Stripe82 in our co-addition
is observed $\sim$50 times.

\section{Reduction}

\subsection{Field layout}

We process the Stripe 82 data in patches of 0.25 square degrees,
covering 0.5 degrees both in right ascension and declination.  Each
declination interval of Stripe 82 is covered by five bins, centred at
Dec. = --1\degr, --0.\degr5, 0\degr, 0.\degr5, 1\degr. The
numbering scheme in right ascension starts in the western part of the
survey, with the first fields centred at RA = 310.\degr25. With
this general layout we assign each field a unique identifier {\sf
  fxxxy}, where the first three digits mark the running number of the
field in right ascension (with {\sf xxx}=001 corresponding to a field
centre at RA = 310.\degr25), using a step size of 0.5 deg
between different fields. The last digit refers to the declination bin
with the declination moving from --1\degr to +1\degr in 0.5 deg
steps when moving from {\sf y}=1 to {\sf y}=5. Field f0164 therefore refers to
a field centre at (RA, Dec.) = (317.\degr75, 0.\degr5).

Data are downloaded from the SDSS DAS using the
{\tt sql} service {\tt casjobs} to select all available images in the
selected region. To avoid boundary effects, we chose a large enough
search radius of 20 arcmin around each field centre in the
selection. Data from the northern and southern strips of Stripe 82 are
not processed separately but combined using the {\tt fpC} images of
all available runs for our co-adds. These {\em `corrected'} images have
been bias corrected, flatfielded and largely removed from instrumental
artefacts. A soft bias of 1000 counts has been added to allow the
archival storage as unsigned integer. All {\tt fpC} images have been
astrometrically calibrated within SDSS, with an accuracy 
better than 45 mas rms \citep{Pier2003}.  Each {\tt fpC} image within
the Stripe 82 data base has a {\tt tsField} fits binary table
associated with it, containing amongst other metadata, calibration
parameters and information on the width of the PSF in the images.

\subsection{Photometric calibration}

We use the information contained in the {\tt tsField} tables for the
photometric calibration of the corrected {\tt fpC} images. The
values of the photometric zeropoint (`$aa$'), extinction coefficient
(`$kk$') and airmass (`airmass') were extracted for all five bands,
yielding the photometric calibration of the {\tt fpC} images according to

\begin{equation} 
\frac{f}{f_0} = \frac{DN}{t_{exp}} \times 10^{0.4\times (aa + kk \times airmass)}
\end{equation}

\noindent with $f$ being the flux in the image, $DN$ its corresponding count
rate and $f_0$ the zero-point flux. Pogson magnitudes in the AB
system are then calculated by

\begin{equation} 
mag = -2.5 \times \log{\frac{f}{f_0}}
\end{equation}

\noindent Before deriving appropriate values for the sky brightness and
corresponding gradients in each image, the {\tt fpC} images have to be
aligned photometrically. This is achieved by correcting each image for
the effects of atmospheric extinction and setting all images to a
common zero-point of $aa'$=--24.0~mag, irrespective of observation date or
waveband.  After subtraction of the soft bias of 1000 counts, we multiply
each image by a correction factor $c$ defined as

\begin{equation} 
c = 10^{0.4\times [(aa + 24) + kk \times airmass]}
\end{equation}

\noindent For the fluxes $f'$=$c \times f$ and count rates $DN'$=$c
\times DN$ in the corrected, photometrically aligned frames the
calibration equations take the following form:

\begin{equation} 
\frac{f'}{f_0} = \frac{DN'}{t_{exp}} \times 10^{0.4\times aa'}
\end{equation}

\noindent and 

\begin{equation} 
mag = -2.5 \times \log{\frac{f'}{f_0}}
\end{equation}

\noindent with a unique zeropoint $aa'$=-24.0~mag.

\subsection{PSF information}

The SDSS photometric pipeline PHOTO fits various profiles for point
and extended sources to objects detected in the {\tt fpC} frames. All
this information is used to obtain values for the shape and width of
the PSF and its corresponding errors. As we use the PSF information
primarily to discard certain images with large PSF widths from the
individual stacks, we extract the PSF information from the {\tt
  tsField} calibration tables (parameter `psf\_width').

\subsection{Sky subtraction}
\label{sec.red_sky}

To reach the very low surface brightness values we aim to achieve with
our Stripe 82 co-adds, we have to determine the sky background in the
single images to a high accuracy. Having measured sky values and their
dispersion in the images, we are then able to discard images with
large background values -- taken under bad observing conditions with
bad sky transparencies or during grey to bright time. In addition, 
we reject images with large variations of the sky background from 
the stacks. To get rid of point and extended sources which would 
affect the calculation of the sky background, we use {\tt SExtractor}
\citep{Bertin1996} to create object masks which include the background
map subtracted count rates of all objects detected in the {\tt SExtractor}
run. By applying a growth radius of 5 pixels to each pixel masked as
object, we make sure that the faint wings of sources do not affect the
estimation of the sky background. Finally, the soft bias-subtracted
and flux-corrected {\tt fpC$'$} images and corresponding object masks
(transferred to binary images comprised of pixels with values equal 1
`=sky' and 0 `=object') are combined, yielding the input frames for
the determination of the sky background. We obtain the sky values by
placing $10^5$ square apertures with side lengths of 21 pixels (or 8.3
arcsec) randomly in the images and measuring the median of the
count rates in each box. The size of the apertures is chosen large
enough to allow a robust measure of the median, but 
small enough to be affected by gradients in the sky
background. $\kappa$-$\sigma$ clipping of these $10^5$ values yields a
first determination of the sky value in the image and its
dispersion. By repeating this procedure five times and averaging over the
results, we finally obtain a very robust measurement of the sky
background $B$ in the images. Depending on the observing conditions,
the background in the images is not completely flat but can show
gradients in both the scanning (RA) and perpendicular (Dec.)
direction. However, as our main intent is on low surface brightness
features, we subtract just a single value for the sky background and do
not try to model it in 1D or 2D. In this way, we preserve as much as
possible the characteristics of the low surface brightness
objects. This is one of the major differences to previous reductions
of Stripe~82 by \citet{Annis2014} and \citet{Jiang2014} which both
apply background modelling in their reduction pipelines.  For larger
gradients and high background values, we apply several selection cuts
which eliminate the data from the final stack and ensure that the data
used in the co-addition have well-defined sky backgrounds. 

\begin{figure}
\centering
\includegraphics[width=0.5\textwidth]{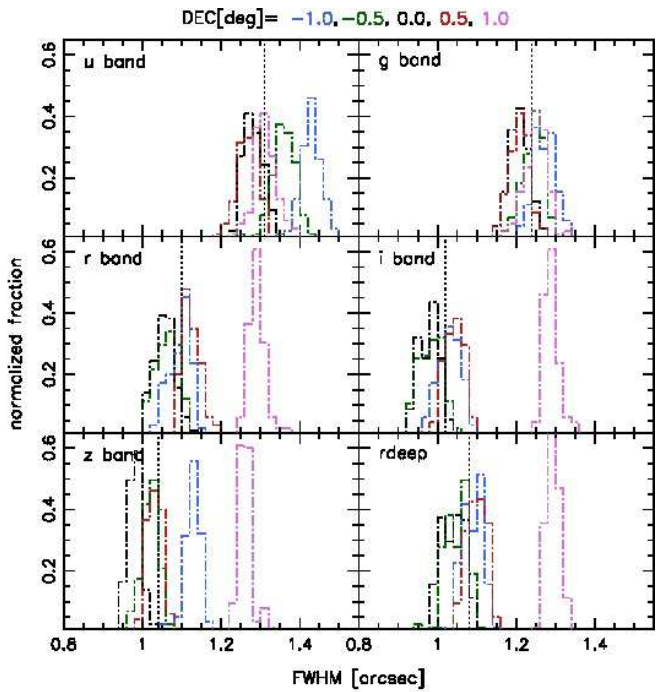}
\caption{Distribution of PSF FWHM values along the Stripe 82
  co-adds. The FWHM shows a strong dependence on the declination of the
  field centre. The median values in each band are indicated by
  vertical dashed lines.
 \label{fig.psf_hist}}
\end{figure}

\subsection{Co-addition}
\label{coaddition}

Images with large sky backgrounds, notable sky gradients, large PSF
widths, but also observations affected by clouds or
bad transparency will degrade the quality and reduce the achievable
depths of the co-added images. We therefore apply a couple of selection
criteria to the images entering the stack. These criteria are based on
the information gathered from the calibration tables (the width of the
PSF and throughput information collected in the flux correction
parameter $c$) and on information resulting from the sky determination
(sky background $B$ and its dispersion $\sigma_B$). Using
the information from all the images, we calculate the $\kappa$-$\sigma$
clipped mean and dispersion of the PSF ($\overline{PSF}$,
$\sigma_{PSF,s}$) and sky ($\overline{B}$, $\sigma_{B,s}$) in the
image samples. For all the frames taken during the photometric runs until
2004, we also calculate the $\kappa$-$\sigma$ clipped mean of the flux
correction factors $\overline{c}$, and impose the following selection
criteria for images entering the stack:

\begin{itemize}

\item $ B < \overline{B} + 2 \times  \sigma_{B,s}$
\item $ PSF < \overline{PSF} + 2 \times  \sigma_{PSF,s}$
\item $ \sigma_B  < \sigma_{lim}$
\item $ c < 1.30 \times  \overline{c}$

\end{itemize}

\noindent with $\sigma_{lim}=(0.5,1.0,1.2,1.2,0.5) \times
\overline{c}$ for $(u,g,r,i,z)$. This last criterion ensures that
images whose throughput values deviate by more than 30\% from the mean
of the photometric runs are discarded from the final stack. In this
way, we eliminate non-photometric data taken under bad observing
conditions. The criteria eliminate on average one third of the data
with the fraction of bad data decreasing for redder wavebands.
The median fractions of eliminated bad data are (39,41,34,26,23) per
cent in $(u,g,r,i,z)$, roughly constant over the survey area. This
means that compared to \citet{Jiang2014}, we only use $\sim$ 80 per cent
of the data which passed their quality criteria for our co-adds.

\begin{table}
\centering
\begin{tabular}{cc}
Band & FWHM (arcsec) \\
\hline
$u$  &  1.31 \\
$g$  &  1.24 \\
$r$  &  1.10 \\
$i$  &  1.02 \\
$z$  &  1.04 \\
\hline
$r_{deep}$ & 1.08 \\
\hline
\end{tabular}
\caption{Median values of the PSF FWHM of the co-added data. \label{tab.psf_median}}
\end{table}

All sky-subtracted and photometrically aligned images which pass the
criteria are fed to {\tt SWarp} \citep{Bertin2002} for co-addition. {\tt SWarp}
takes the information on the World Coordinate System provided in the
image header and projects the images on a regular grid with a pixel
size of 0.396 arcsec before co-adding the images on a pixel per pixel basis.
As we have already subtracted the sky from the images, we run {\tt SWarp}
without the sky subtraction option, choosing Lanczos3 as interpolation
kernel during the regridding process which produces a tangent plane projection
centred on the field centres.  As output, {\tt SWarp} yields the co-added
image, calculated as unweighted median of the pixel values of all
selected images, and a weight image providing the number of input
frames contributing to the final stack on a pixel-per-pixel basis.
The median ensures a robust removal of outliers and image defects.

\begin{figure*}
\centering
\includegraphics[width=0.49\textwidth]{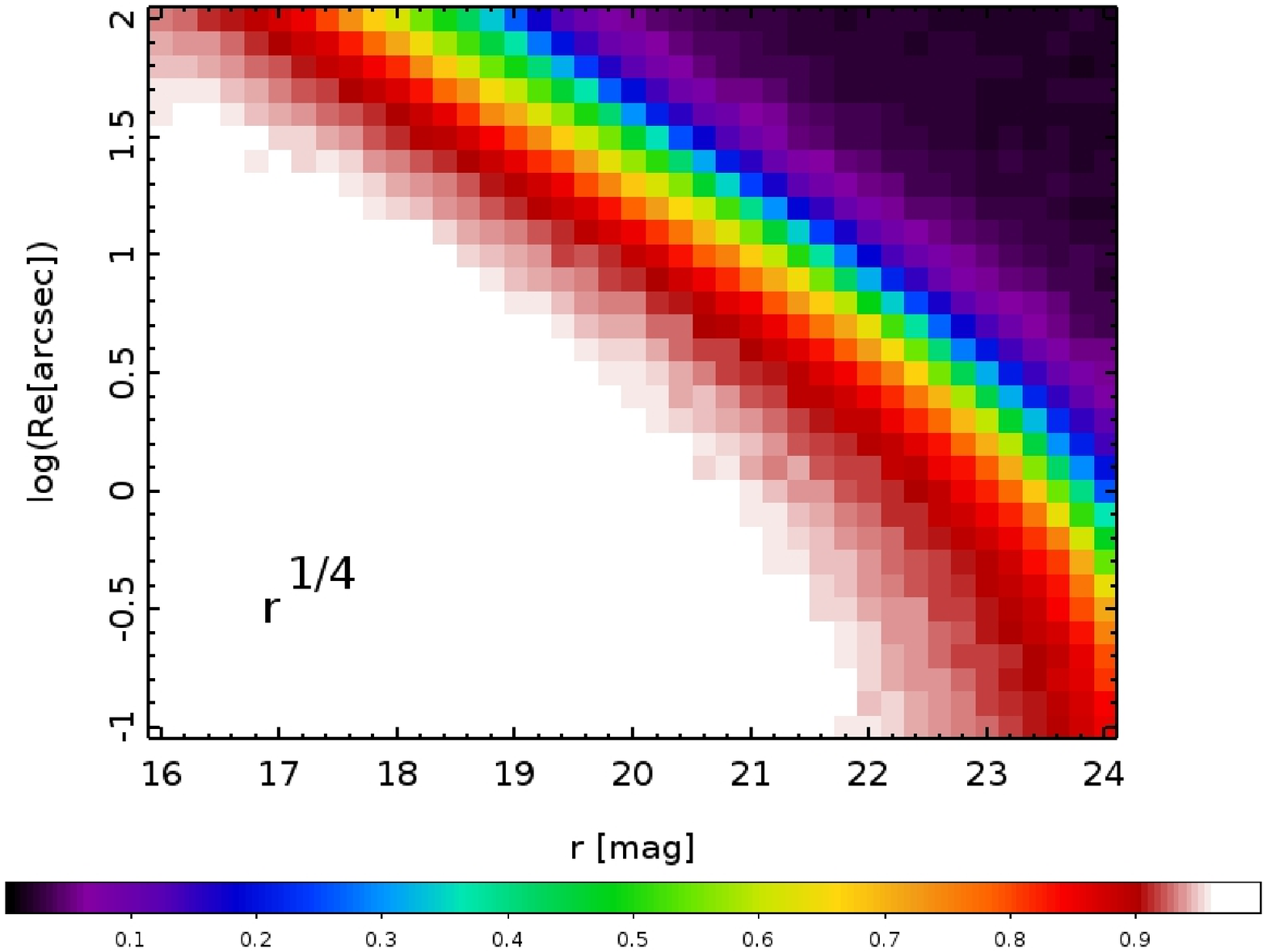}
\includegraphics[width=0.49\textwidth]{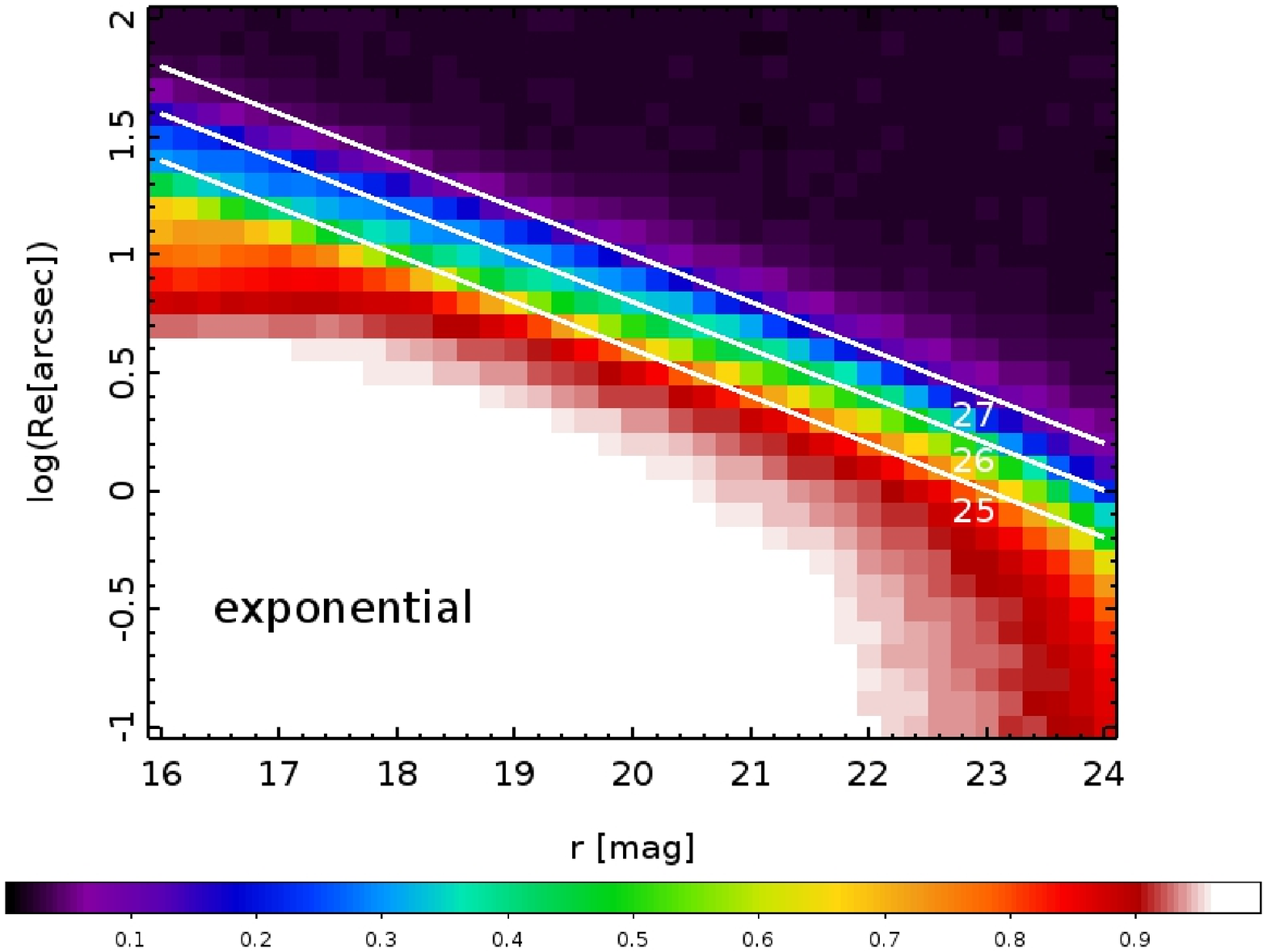}
\caption{Detection map (recovery fraction as a function of input
  magnitude and effective radius) in the $r$ band for simulated de
  Vaucouleur (left-hand panel) and exponential (right-hand panel) profiles with
  $\log(R_E$[arcsec]) ranging from --1.0 and +2.0. For the exponential
  profiles, we include the lines of equal effective surface brightness
  $<\mu_e>$=25, 26, 27 mag~arcsec$^{-2}$ (from bottom to top). The
  50\% completeness level lies at $<\mu_e ($r$)>$ $\sim$25.5~mag
  arcsec$^{-2}$. \label{fig.completeness_ext}}
\end{figure*}

\section{Data quality}
\label{sec.quality}

\subsection{PSF values}
\label{sec.quality_psf}

We determined full width at half maximum (FWHM) values of the PSF for all
our co-adds using the {\tt PSFex} software \citep{Bertin2011}. Image
representations of the calculated PSFs are included in the data
release (see Sec.~\ref{sec.release_psf}). They can be plugged in
easily to be used e.g. for model convolution within {\tt GALFIT}
\citep{Peng2002}, {\tt IMFIT} \citep{Erwin2015} or similar
applications. {\tt PSFex} relies on {\tt SExtractor} catalogues in combination
with small vignettes of detected objects in the images. Point sources
are selected automatically by their determined half-light radius and
brightness properties. We omitted any fitting or oversampling of the
PSF to obtain its natural shape in the SDSS Stripe 82 fields, using
the information of the whole 0.25 square degree co-add
for the creation of deep PSFs in all bands.

In Table~\ref{tab.psf_median} we list the median values of the
FWHM. As expected the PSF is narrower in the redder bands with $r$,
$i$, and $z$ showing values below 1.1~arcsec. The combination of the
$g$, $r$ and $i$ bands (what we call $r_{deep}$) has a value of 
1.08~arcsec. The $u$ band has the broadest PSF. Still, with 1.3~arcsec, 
its value is sufficiently
small for most ground-based astronomical studies.  However, there is a
strong dependence of the median width of the PSFs on the declination of
the field centres, the dependence on right ascension is only mild (see
Fig.~\ref{fig.psf_hist}). The fields centred at Dec.=+1\degr show
similar PSF widths in all bands which could be up to 30 per cent larger than
the respective median values. In these off-centre fields close to the
northern limit of Stripe~82, the optics determine the PSF width. Also
the fields at the opposite side of Stripe~82, centred at
Dec.=--1\degr, are affected by the optics, albeit to a smaller degree
than the northern fields.  Here only the PSFs in the $u$ and $z$ bands
show distributions shifted towards significantly larger values than
the ones of the central fields close to Dec.=0\degr.

\subsection{Photometric accuracy}
\label{sec.quality_photometry}

We investigated the accuracy of our photometric calibration by
comparing aperture photometry of stars in our source catalogues with
the SDSS standard star catalogue for Stripe 82 \citep{Ivezic2007}. The
source catalogues are based on {\tt SExtractor} photometry on our co-added
images. They are described in more detail in Sec.~\ref{sec.dr_cat}.
For the comparison, we used aperture magnitudes derived in fixed
apertures of 30 pixels (11.88 arcsec) and required clean photometry
flags which eliminate blended objects. From the standard star
catalogue we selected stars with photometric errors below 0.01 mag,
i.e. stars brighter than (19.5, 20.5, 20.5, 20.0, 18.5)~mag in
($u$,$g$,$r$,$i$,$z$). The quoted spatial variation of its photometric
zero-points is not larger than $\sim$0.01~mag (rms). For each of the
co-added images, we cross-correlated the catalogues with a search radius
of 1 arcsec and fitted Gaussians to the magnitude differences. The
zero-point offsets determined in this way were then applied
individually to the co-added images. Median values of the corrections
were (0.03, 0.02, 0.02, 0.03, 0.03)~mag in ($u$,$g$,$r$,$i$,$z$), i.e they
range between two and three times the zero-point accuracy of the standard
star catalogue. These small zero-point variations can be ascribed to
the accuracy of the calibration information provided for each {\tt
  fpC} input image and the inclusion of images whose throughput
factors deviate up to 30 per cent from the mean of the photometric runs (see
Sec.~\ref{coaddition}).  

\subsection{Completeness simulations}
\label{sec.quality_complete}

To estimate the depth of our co-added images we performed completeness
simulations, adding point sources and extended sources of different
profile shapes and effective radii to the co-added images and deriving
the recovery fraction.  We generated deep PSFs using the {\tt PSFex}
software package which represent the input objects for the point
source simulation. PSF convolved frames of the extended sources were
constructed using {\tt GALFIT}. To avoid crowding effects, single simulation
runs consisted in adding between 250 and 500 objects to the images
depending on the effective size of the objects. These runs were
repeated multiple times to a total number of 5000 objects being
injected in the images per filter and size-brightness combination.

Input objects were detected and measured by {\tt SExtractor}, requiring
3$\sigma$ detections with 3 connected pixels lying a factor of 1.74
above the rms value of the background for each pixel. For the
background determination we used a mesh size of 20 pixels and a
background filter of 3 pixels width. Input objects were counted as
detected if input coordinate and the {\tt SExtractor} detection agreed
within 1 arcsec.

Figure~\ref{fig.completeness_ext} shows the detection maps (recovery
fraction as a function of input magnitude and effective radius) for
extended sources with deVaucouleur and exponential surface brightness
profiles and effective radii ranging from $\log(R_E$[arcsec])=--1.0 to
$\log(R_E$[arcsec])=+2.0. For the exponential profiles we include the
lines of equal effective surface brightness $<$$\mu_e$$>$, calculated
as

\begin{equation} 
<\mu_e> = m_T + 5 \times \log(R_E) + 1.995
\end{equation}

\noindent for galaxies with a total luminosity $m_T$ and an effective
radius $R_E$. As Fig.~\ref{fig.completeness_ext} shows, we reach
$<$$\mu_e$$>$$\sim$25.5~mag arcsec$^{-2}$ at the 50 per cent completeness level in
the $r$ band. The 50\% completeness values for the other bands are
(25, 26, 25, 24)~mag arcsec$^{-2}$ in $(u,g,i,z)$.

\begin{figure*}
\centering
\includegraphics[width=0.49\textwidth]{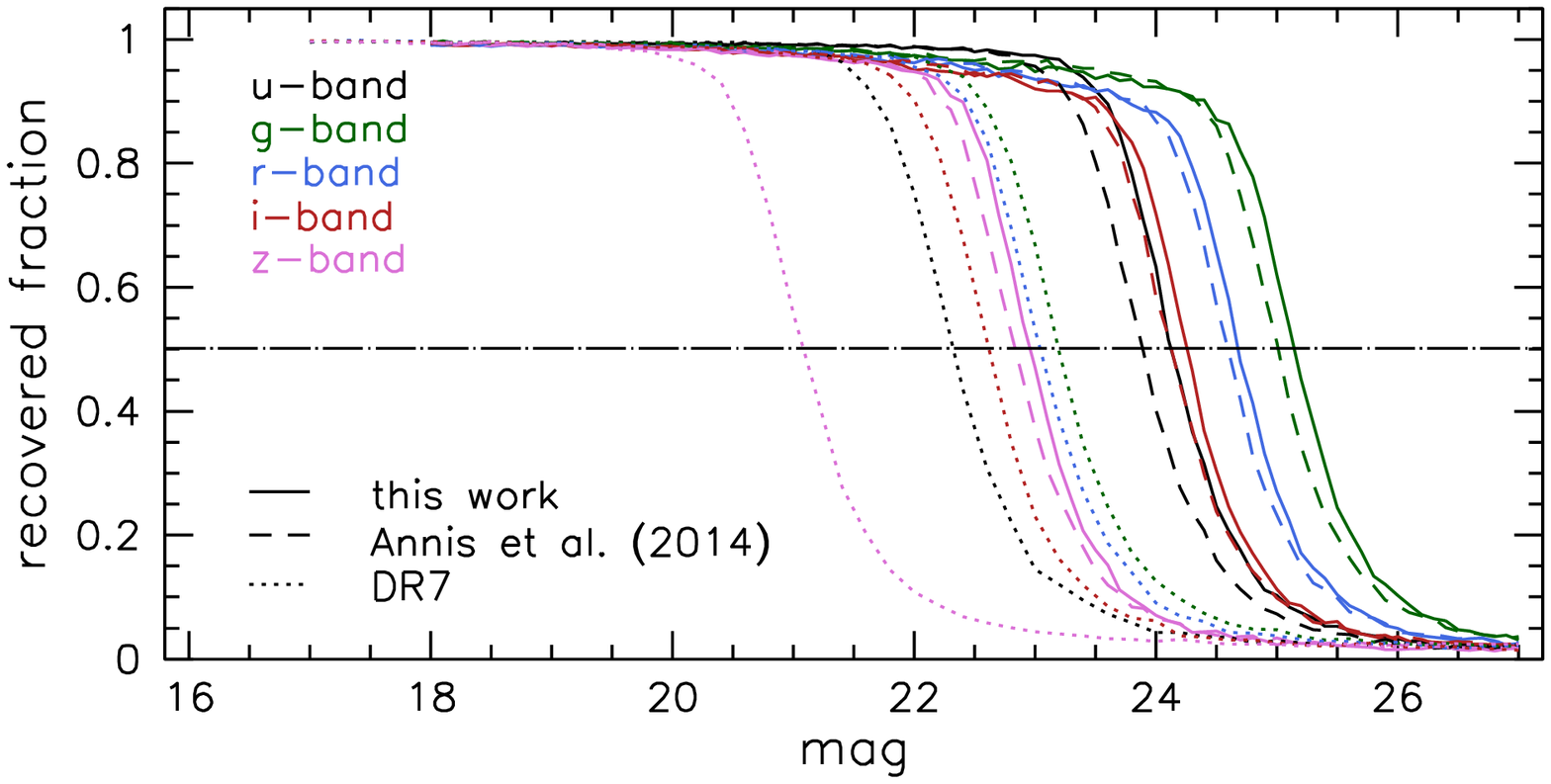}
\includegraphics[width=0.49\textwidth]{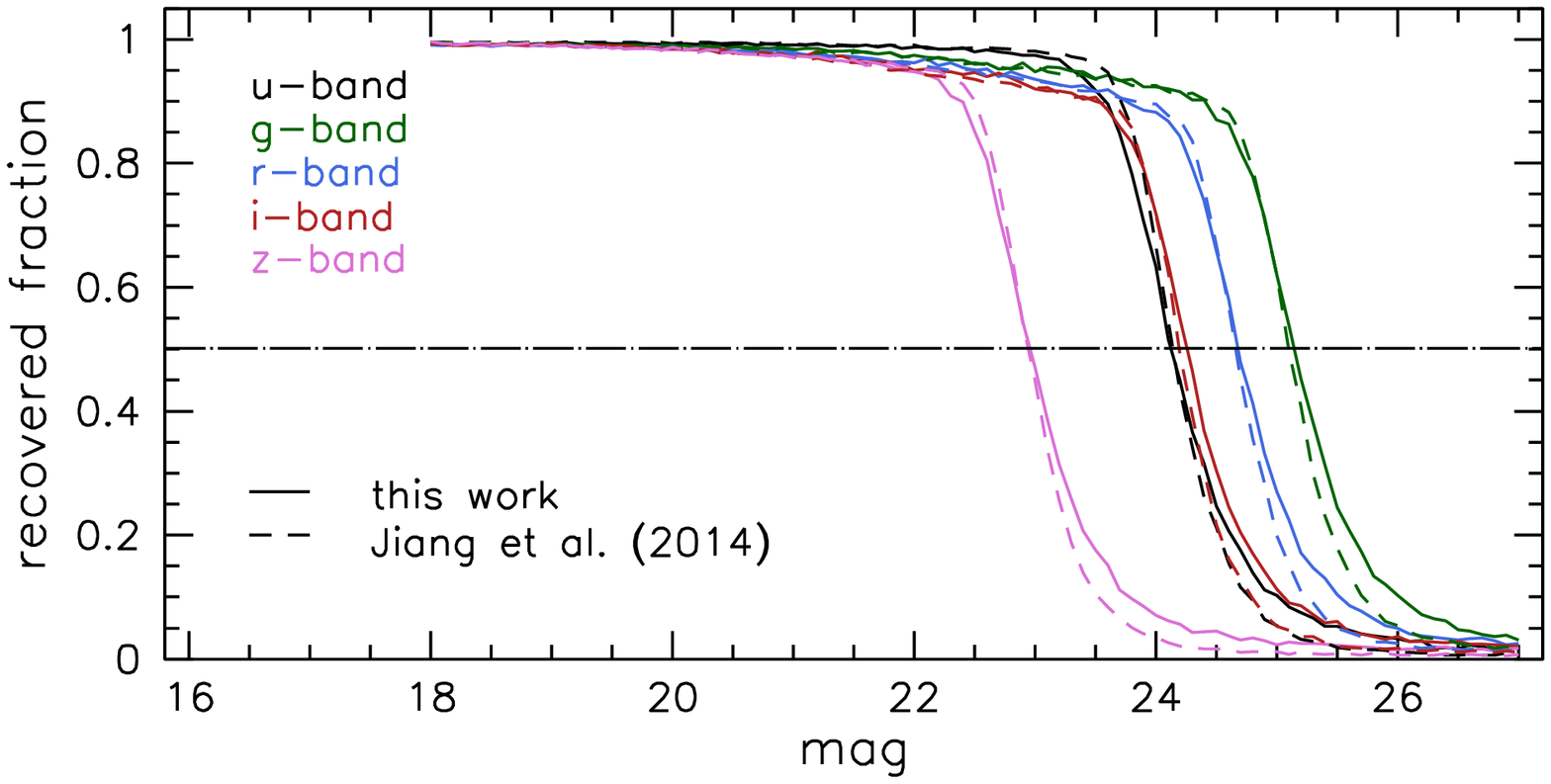}
\caption{Left-hand panel: recovery fraction of point sources as
  a function of input magnitude derived in a typical field of the
  Stripe 82 co-adds. The solid lines show the completeness curves for
  the IAC Stripe 82 co-adds. The comparison with the \citet{Annis2014}
  data is shown as dashed lines. The IAC reduction reaches between 0.1
  mag and 0.3 mag deeper than the \citet{Annis2014} co-adds. Compared
  to the single-epoch DR7 release (dotted lines), the gain in depth
  for the detection of points sources ranges between 1.7 and 2.0 mag.
  Right-hand panel: recovery fraction of point sources as a
  function of input magnitude. The solid lines show the completeness
  curves for the IAC Stripe 82 co-adds; the comparison with the
  \citet{Jiang2014} data is shown as dashed lines. The depth of both
  reductions is similar in all bands.
\label{fig.completeness_ps}}
\end{figure*}

We compared the depth of the IAC co-adds with the Stripe 82 co-adds of
the same area of \citet{Annis2014}, available as runs 10006 and 20006
from the SDSS data archive, as well as with the SDSS DR7 single-epoch
images. Both the Annis et al. co-adds and the DR7 data were aligned
photometrically to the IAC co-adds using the information provided in
the {\tt tsField} calibration tables. PSFs have been created in the
same way as for the IAC co-adds.
Fig.~\ref{fig.completeness_ps} shows the recovery fraction of point
sources as a function of input magnitude, derived in a typical field of
the Stripe 82 co-adds. The gain in depth relative to the single epoch
DR7 observations is around 1.7 magnitudes in $r$ and $i$ and 1.9--2.0
magnitudes in the $u$, $g$ and $z$ bands. In the $g$ band, we reach a
50\% completeness limit of 25.2~mag.  Compared to the
\citet{Annis2014} data, the IAC co-adds reach deeper; the
gain in depth ranges from 0.1~mag in the $r$, $i$ and $z$ bands to
0.3~mag in the $u$ band. The $g$ band lies between these values with
a gain of 0.2~mag (see Fig.~\ref{fig.completeness_ps}, left-hand panel).

For the comparison with the \citet{Jiang2014} data, we proceeded
slightly differently. As their reduction included a 2D modelling of
the sky background by fitting a hyperplane to the data, the
rms in their images and hence the detection threshold are biased
towards smaller count rates, making a comparison of the recovery rate
difficult. We therefore decided to apply the $3\sigma$ threshold as
determined by {\tt SExtractor} for the IAC co-adds also for the {\tt SExtractor}
runs on the \citet{Jiang2014} co-adds which allows a one-to-one
comparison of the detection efficiency. For these completeness
simulations we again aligned the \citet{Jiang2014} data
photometrically to the IAC co-adds using the provided zero-points and
created PSFs in the same way as for the IAC co-adds.

The depth of the IAC and \citet{Jiang2014} co-adds is similar in all
bands, i.e. 0.1--0.3 mag deeper than the \citet{Annis2014} data (see
Fig.~\ref{fig.completeness_ps}, right-hand panel). Interestingly, we did
not notice the 0.3--0.5 mag difference in depth between
\citet{Jiang2014} and \citet{Annis2014} as reported in
\citet{Jiang2014}, derived as the magnitude at which the magnitude
error exceeds 0.22~mag. We list our determined 50\% completeness
values of all three Stripe 82 reductions and the DR7 single epoch data
in Table~\ref{tab.completeness}. 

\begin{figure*}
\centering
\includegraphics[width=0.49\textwidth]{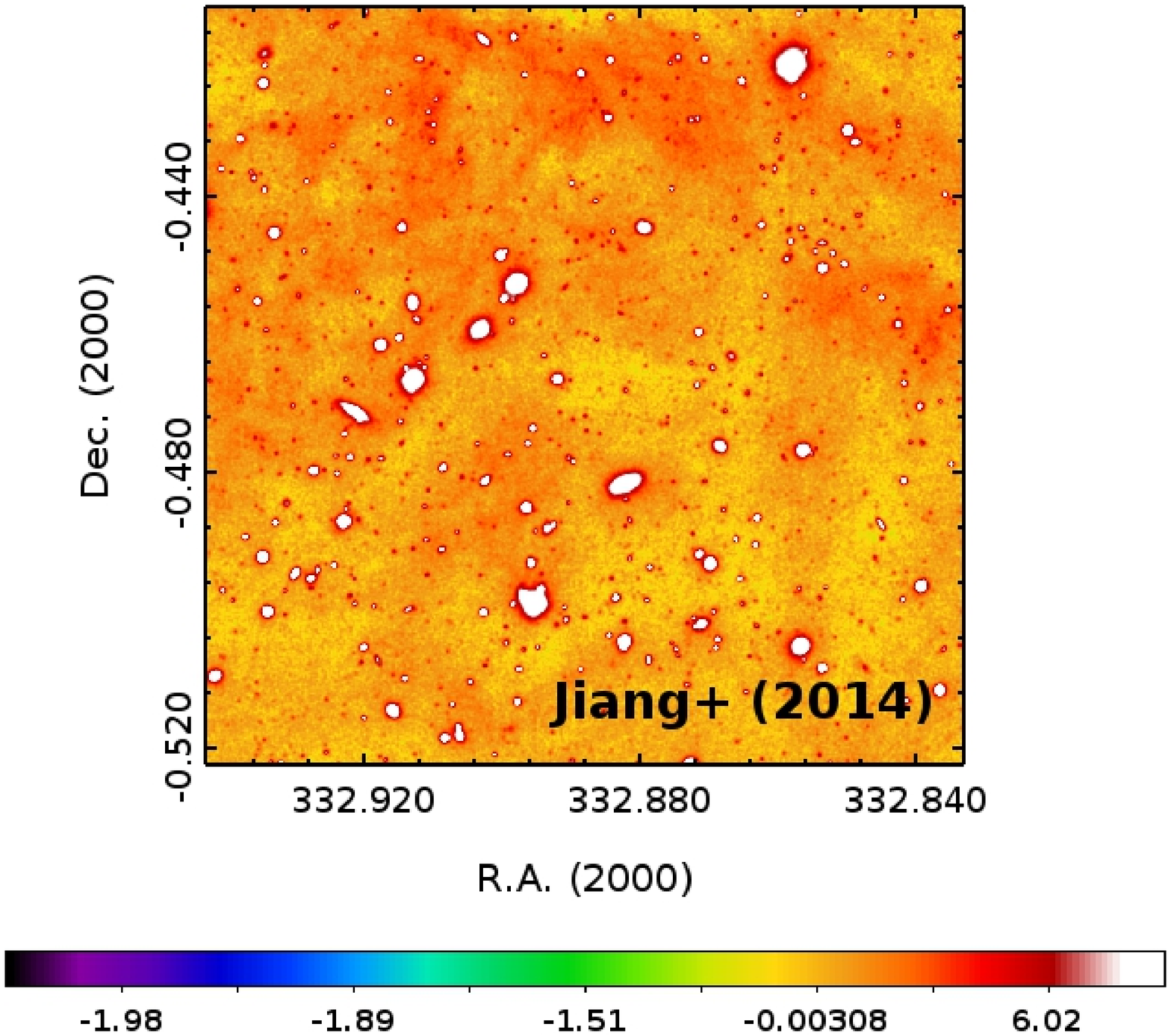}
\includegraphics[width=0.49\textwidth]{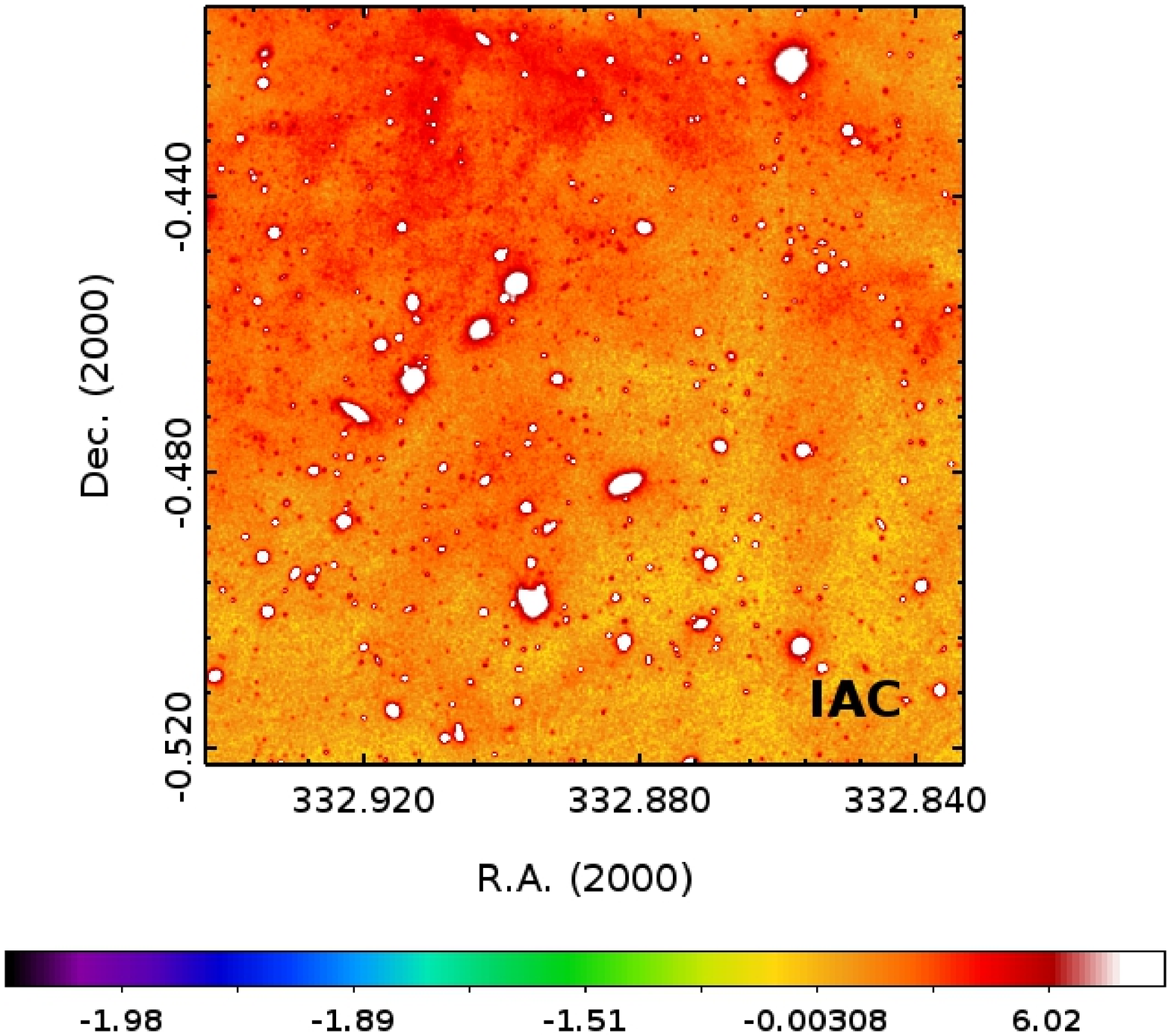}
\hspace*{0.1cm}\includegraphics[width=0.33\textwidth]{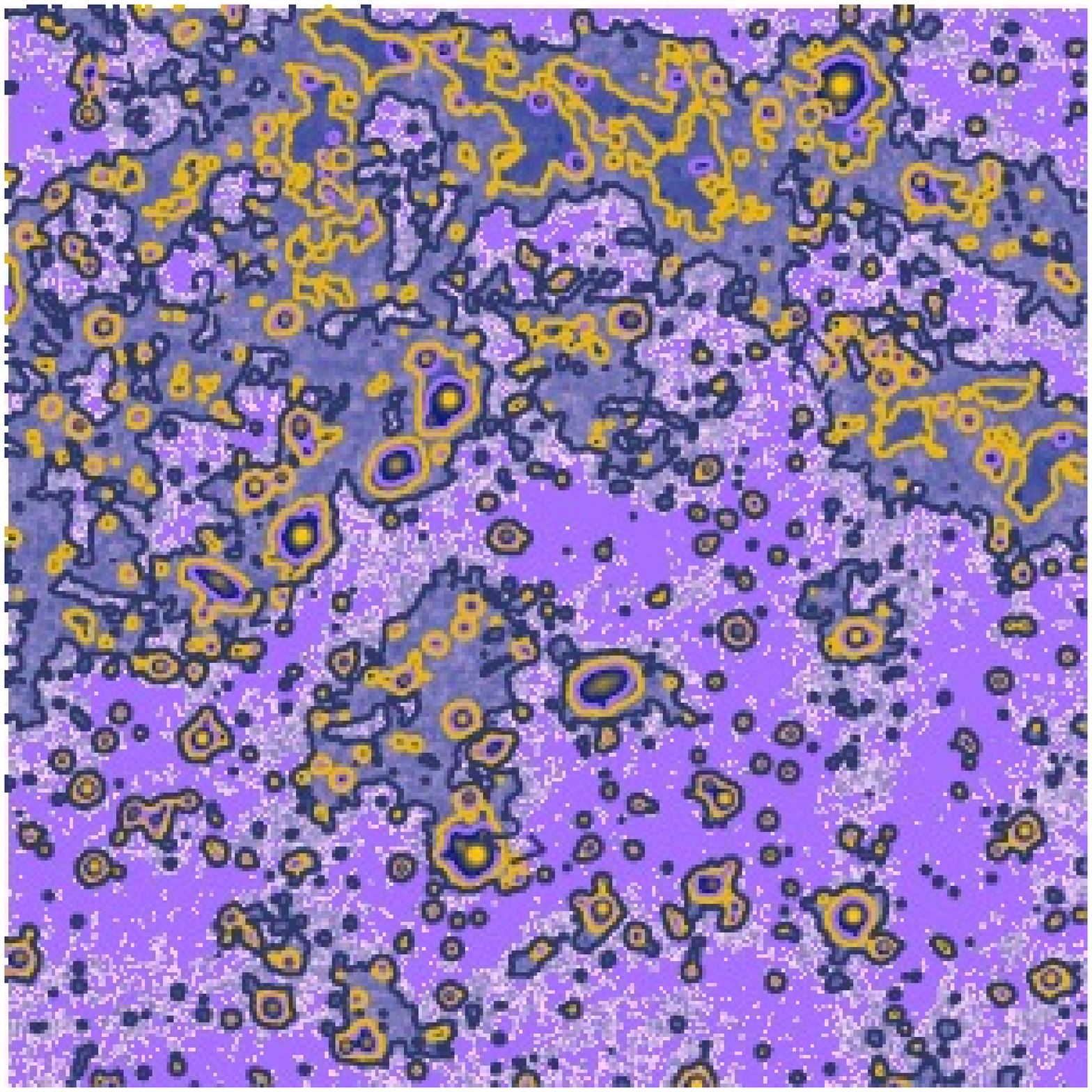}
\hspace*{2.9cm}\includegraphics[width=0.33\textwidth]{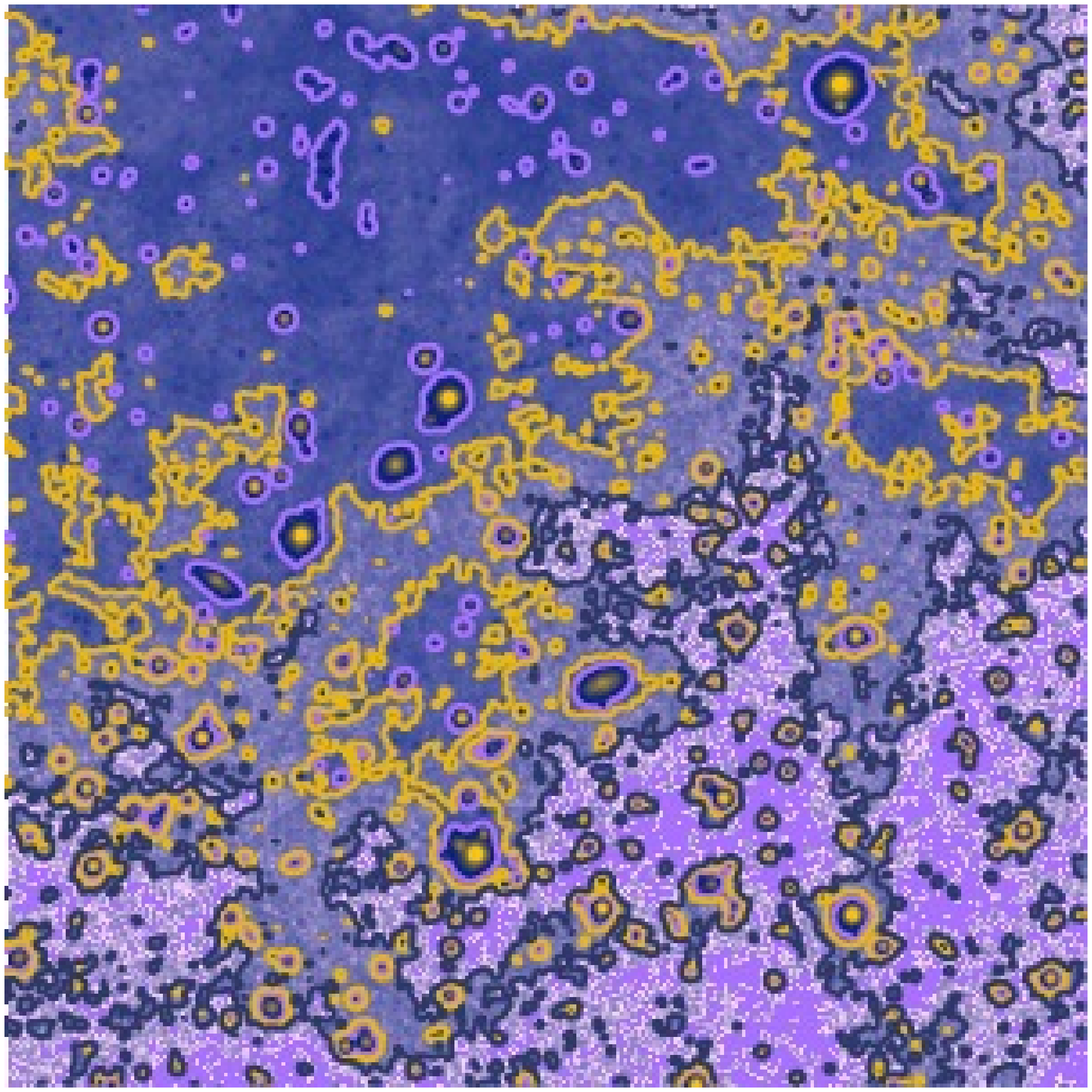}
\caption{Comparison of a field showing intense emission of Galactic
  Cirrus in \citet{Jiang2014} (left-hand panels) and our data (right-hand
  panels). The top panels show the co-adds in the $r$ band in the same
  dynamic range. Both images were photometrically aligned to the same
  zeropoint. Additionally, we applied corrections of +0.16 and +0.34
  $DN$ to the \citet{Jiang2014} and our image, respectively. These
  values correct for a slight negative offset of the background in
  pure sky regions. They have been determined as the median of the
  count rates in selected empty regions in the field.  The bottom
  panels show the surface brightness distribution of the Galactic
  Cirrus in both images. Contour values are 25, 26 and 27
  mag~arcsec$^{-2}$. Compared to our images, the cirrus emission is
  largely reduced in the \citet{Jiang2014} data, presumably due to
  differences in the background treatment in the reduction process.
  \label{fig.comparison_dust}
}
\end{figure*}

Simulations of extended objects follow the same trend as shown for the
point sources, i.e. the data of \citet{Annis2014} being slightly
shallower than the IAC and \citet{Jiang2014} data. Note, however, that
this comparison gives only a lower limit on the difference in depth
for extended objects between our and previous reductions. As the
simulated objects are ingested in the final co-adds, the results only
reflect the differences in the noise properties between the co-added
images. They do not account however for differences in the reduction
process which could affect the characteristics of extended objects,
especially at the low surface brightness limit.  This is exemplified
in Fig.~\ref{fig.comparison_dust} which shows a field with intense
emission of Galactic Cirrus (see Sec.~\ref{sec.cirrus}) in our
reduction and the reduction by \citet{Jiang2014}. Both images show
large differences in the brightness of the dust clouds. In the regions
with maximum emission, the intensity in the \citet{Jiang2014} data is
reduced by roughly a factor of 2 compared to our data. This
reduction in flux originates presumably from differences in the
treatment of the sky background in the input images. Our reduction
subtracts just a single background value and therefore preserves the
characteristics of the background composed of sky and diffuse light.
\citet{Jiang2014} on the other hand subtract a 2D model of the
background which can potentially misidentify faint astronomical
objects as sky background and eliminate them from the reduced images
or alter their fluxes.

\begin{table}
\centering
\begin{tabular}{cccccc}
Band & DR7 & Annis et al. & Jiang et al. & IAC & Gain \\
\hline
$u$  & 22.3 & 23.9   & 24.2   & 24.2 &  1.9 \\
$g$  & 23.2 & 25.0   & 25.1   & 25.2 &  2.0 \\
$r$  & 23.0 & 24.6   & 24.7   & 24.7 &  1.7 \\
$i$  & 22.6 & 24.2   & 24.2   & 24.3 &  1.7 \\
$z$  & 21.1 & 22.9   & 23.0   & 23.0 &  1.9 \\

\hline
\end{tabular}
\caption{50\% completeness limits for points sources, derived for all
  three reductions of the Stripe 82 data set. For reference, we give
  also the determined values for the single epoch SDSS DR7 release
  and the gain in depth of the IAC co-adds presented in this work.
\label{tab.completeness}}
\end{table}

\begin{figure*}
\centering
\includegraphics[width=0.32\textwidth]{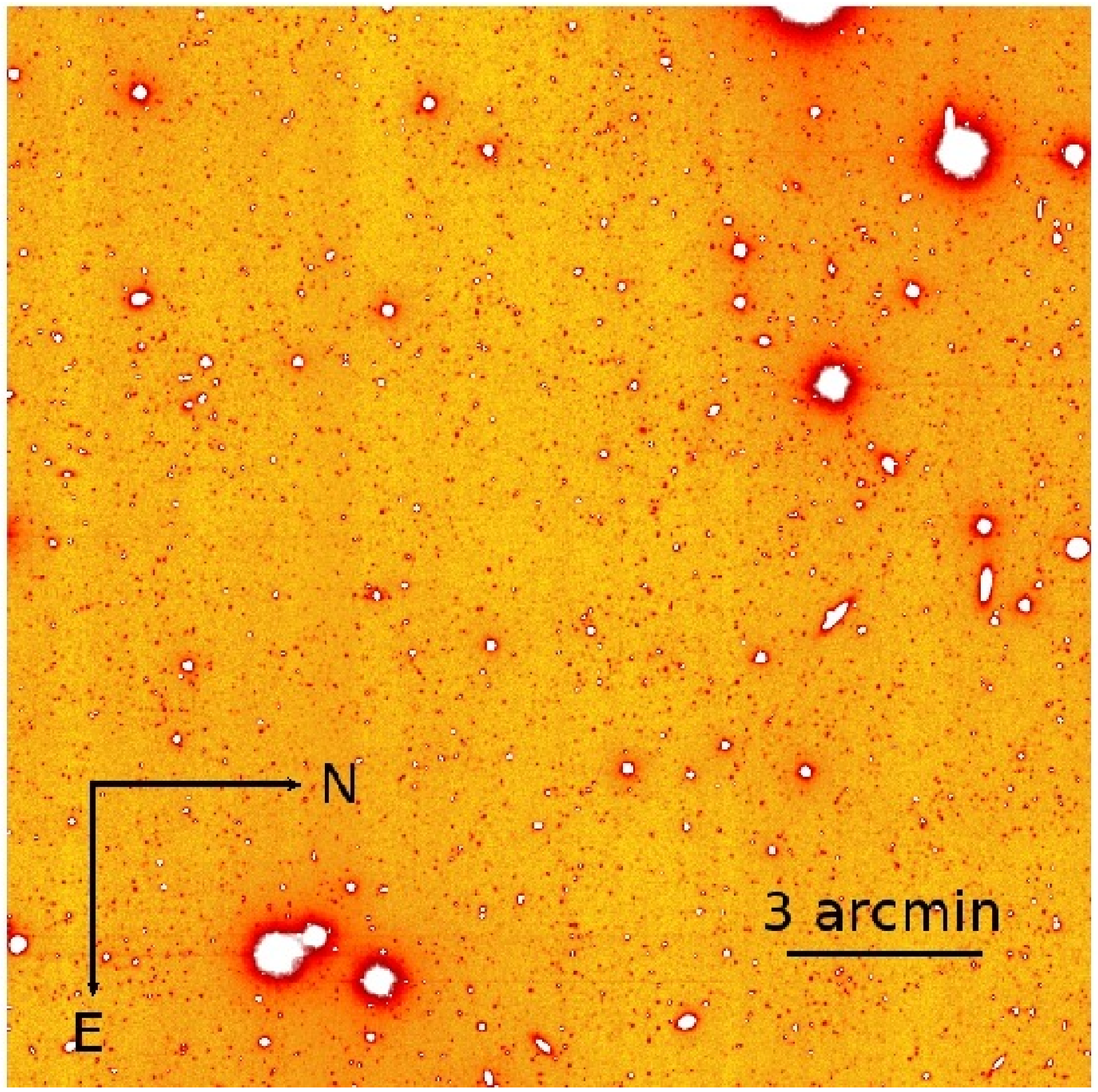}
\includegraphics[width=0.32\textwidth]{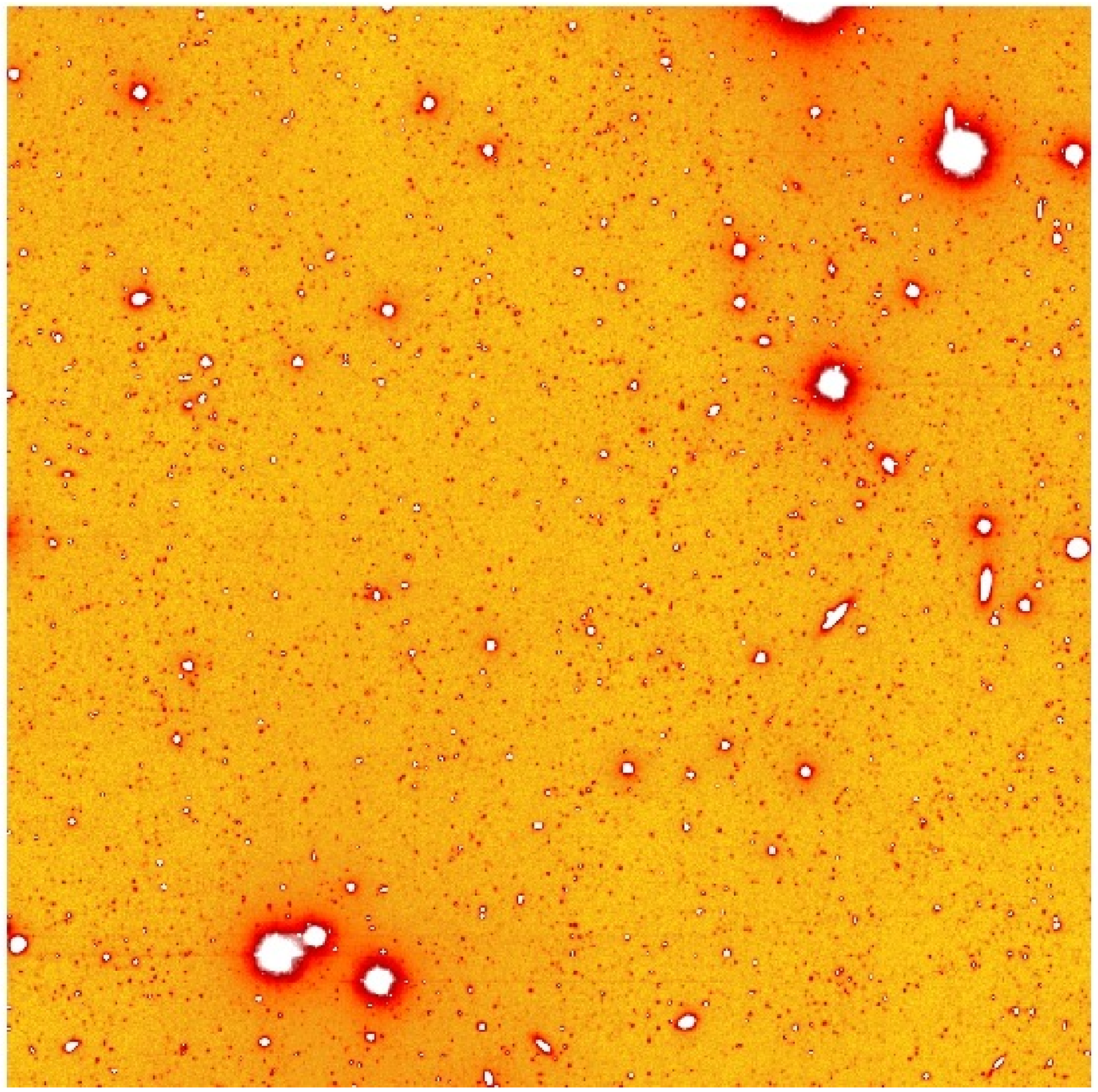}
\includegraphics[width=0.32\textwidth]{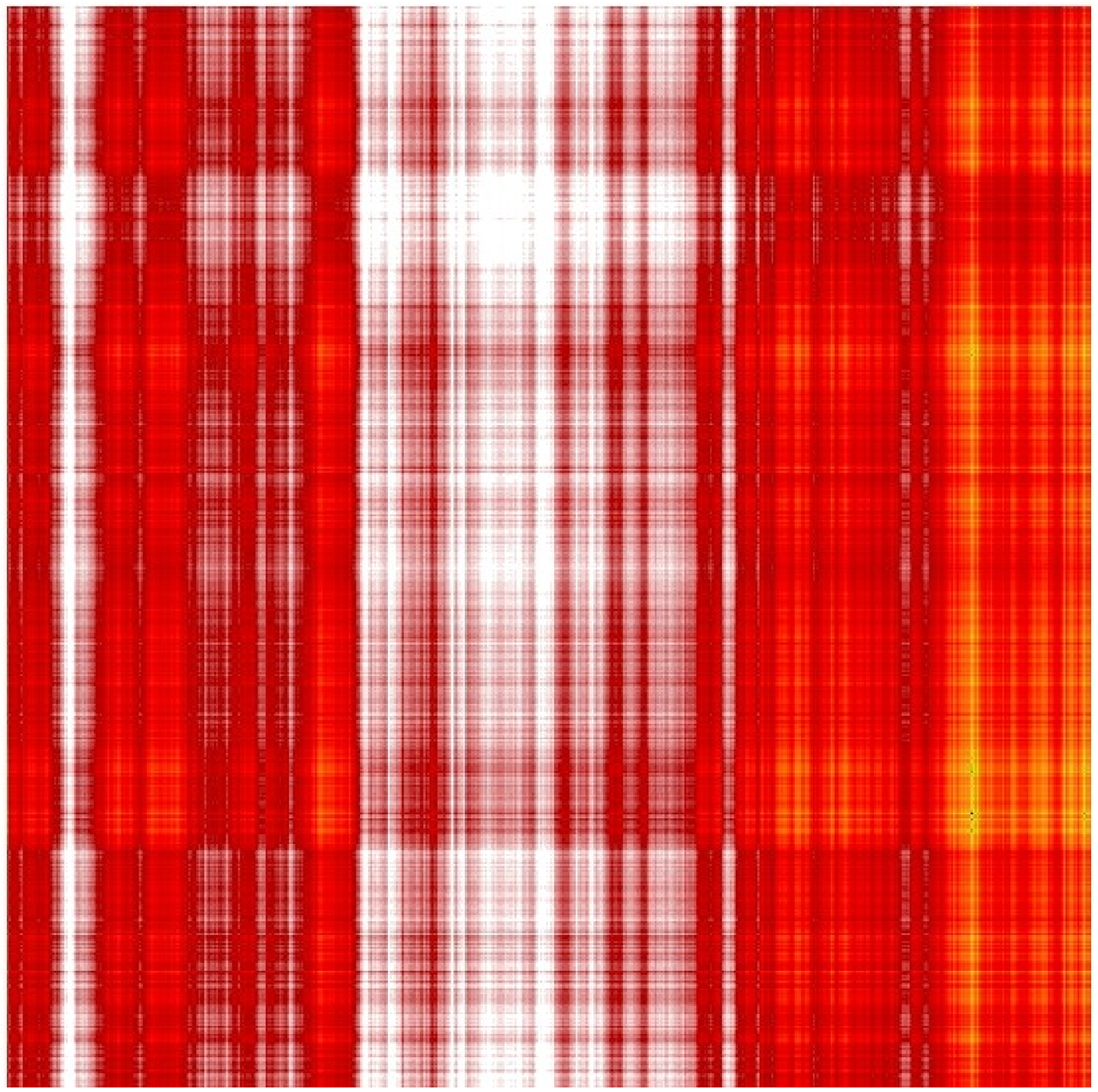}
\caption{Example of the sky rectification method. Left-hand panel:
  the original co-add in $r_{deep}$ shows small gradients along the
  Dec. direction. Middle panel: sky-rectified version of the
  same image. Right panel: correction applied to the data; the
  frame shows the expected pattern of vertical and horizontal
  lines, but no imprints of objects. \label{fig.dr_sky}}
\end{figure*}

\section{Data release}
\label{sec.dr}

The data products of the Stripe 82 reduction are released through a
dedicated webpage at http://www.iac.es/proyecto/stripe82. The release
contains the co-added data in 5+1 bands, sky-rectified versions of the
co-adds, corresponding exposure time maps, image representations of the
PSF for each co-add and catalogues of point and extended sources.

\subsection{Co-adds}
\label{sec.dr_coadd}

The co-added data are provided in patches of 0.5$\times$0.5 square
degrees.  In this way, 5 images cover the full range of declination
(--1.25$\degr$ to +1.25$\degr$) at a certain right ascension in the
Stripe 82 region.  The data are calibrated to a common zero-point of
24~mag for all bands.  AB magnitudes can be derived from the data
via

\begin{equation}
mag = -2.5 \times  \log\big(\frac{DN}{t_{exp}}\big) + 24.0 
\end{equation}

\noindent where $t_{exp}=53.907456$~s refers to the exposure time of a single
SDSS image and $DN$ to the counts measured in the co-add.

Each co-added image has an exposure time map attached to it which
yields for each pixel the number of single images $n(x,y)$ entering the
stack. The effective exposure time in each part of the image is given
by the product $n(x,y) \times t_{exp}$.

Additionally, we provide images representing the average of the $g$,
$r$ and $i$ co-adds which, for a flat object spectral energy distribution, 
can be regarded as
deep $r$-band co-adds. As the calibration gets destroyed during this
process, above equations are invalid for the deep $r$-band
($r_{deep}$) images. Nevertheless, the gain in depth of 0.2--0.3
mag compared to the single co-adds can give valuable
insights into low surface brightness features close to the detection
limit of the survey.

\subsection{Sky-rectified co-adds}
\label{sec.dr_sky}
Small gradients in the background of similar shape and amplitude are
still apparent in the co-adds from \citet{Annis2014} and our work,
predominantly along the Dec. direction. As the reduction of
\citet{Jiang2014} included a 2D modelling of the sky background in the
input SDSS images their co-adds show the smallest gradients. To
preserve as much as possible the intensity and shape of objects at low surface
brightness, we decided to characterize the sky properties of the input
SDSS images by a single value and tackle larger gradients by the
measured variance of the sky background over the image. Remaining
small gradients are treated in the co-adds at much higher
signal-to-noise (S/N) than in the input SDSS images. In this way, the
probability of confusing sky background and astrophysical objects at
low surface brightness is reduced.

\begin{figure}
\centering
\includegraphics[width=0.4\textwidth]{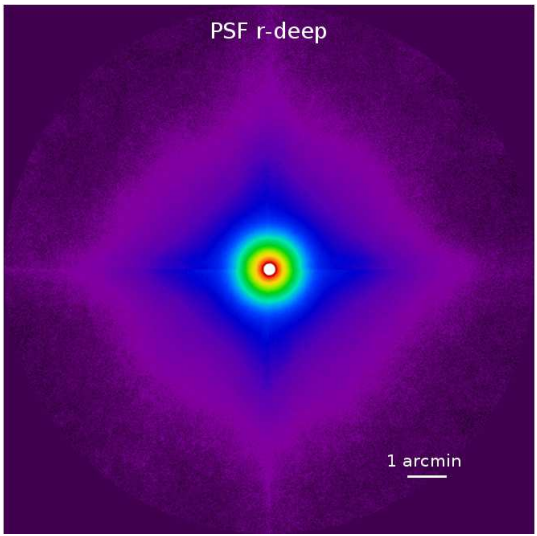}
 \caption{Ultra-deep PSF in the $r_{deep}$ band. The PSF was created
   by combining the PSF of bright saturated stars for the outer part
   with {\tt PSFex} created high-S/N PSF for the inner part. The dynamical
   range of the PSF extends over 20 mag.
\label{fig.dr_psf}}
\end{figure}

For sky-rectified versions of our co-adds, we proceed as follows.
Object masks are created independently for the co-adds in each filter.
Here, we put large emphasis on conservatively masking the haloes of
stars and galaxies but also the areas with Galactic Cirrus or diffuse
low surface brightness emission. The masks also include the extended
PSF wings of bright stars and ghost artefacts. We then calculate the
$\kappa$-$\sigma$ clipped mean of the lines perpendicular to the direction
of the gradient, i.e. the RA direction, and subtract the value from the
count rates in these columns. To eliminate small inhomogeneities in the
perpendicular direction, we calculate the $\kappa$-$\sigma$ clipped mean
of the rows and subtract it; this step uses the already column-corrected
image from the previous step as input. As example resulting from this
method we show in Fig.~\ref{fig.dr_sky} the original
co-add, the sky-rectified version and the difference of both, i.e. the
correction applied to the data. The correction frame shows the
expected pattern of vertical and horizontal lines with no additional
features related to objects in the field of view. Our tests have shown
that the method is flux-conserving and preserves the quantities of low
surface brightness features like Galactic Cirrus.

\subsection{Deep PSFs \label{sec.release_psf}}
\label{sec.dr_psf}

Representations of the PSF are provided for each co-add in the five SDSS
bands plus the $r_{deep}$ images. The PSFs are calculated using the
{\tt PSFex} software which selects point sources from {\tt SExtractor} catalogues,
normalizes and combines them to a robust mean PSF for a particular
field. The selection of point sources is done in an automatic way
using the half light radius and luminosity information included in the
{\tt SExtractor} catalogues. Depending on the filter and the location of the
field within Stripe 82, between 300 and 2000 point sources are used
for the determination of the PSF, with the smallest numbers applying
to the $u$ band, as expected. The rms variation of the FWHM of
the PSF across our 0.25 square degrees fields ranges between 1 per cent
(median value) in the central fields close to the optical axis and
up to 5 per cent (median value) in the northern fields at Dec.=+1\degr,
with the spread being largest in the $i$ band and smallest in the
$u$ and $g$ bands. The rms is mostly dominated by the variation
of the PSF along the Dec. direction, causing generally broader PSFs
in the off-centre fields (see Sec.~\ref{sec.quality_psf} and
Fig.~\ref{fig.psf_hist}). As each portion of the Stripe 82 co-adds is
composed of images sampling the whole range in positions on the CCD
along the scanning (RA)  direction, the PSF dependence on the
RA in the individual co-adds is weak.

The properties of stellar haloes and the faint outskirts of galactic
discs can be largely influenced by the wings of the stellar PSF
\citep[e.g.][]{Zibetti2004,Sandin2014,Trujillo2015}.  To estimate this
effect in the Stripe 82 data, we provide ultra-deep PSFs reaching out
to 8~arcmin in radius and extending over 20 mag in
dynamical range. The ultra-deep PSFs were created by combining the
PSFs of bright saturated stars and {\tt PSFex}-generated PSFs of certain
fields. For the saturated part, we selected 102 bright stars with
6$<$$R$$<$8~mag within the Stripe 82 region from the
United States Naval Observatory B1.0 catalogue (Monet et al. 2003).
The stellar images were aligned to a common grid based on the catalogue
positions and normalized to a common flux by using their catalogue magnitudes:
$B$ was used for the $u$ and $g$ bands, $R$ for the $r$ band, and $I$ for
$i$ and $z$ bands. In every band, the final PSF was obtained by
median-combining each stack of images. For the central saturated part
of the PSF, we use a high S/N PSF created by {\tt PSFex} on a frame covering
several Stripe 82 fields.  After matching the profiles of saturated
and non-saturated PSFs to align the flux scaling, both PSFs were
combined to the ultra-deep PSF. As example, we show the ultra-deep PSF
in $r_{deep}$ in Fig.~\ref{fig.dr_psf}.

\begin{figure}
\includegraphics[width=0.5\textwidth]{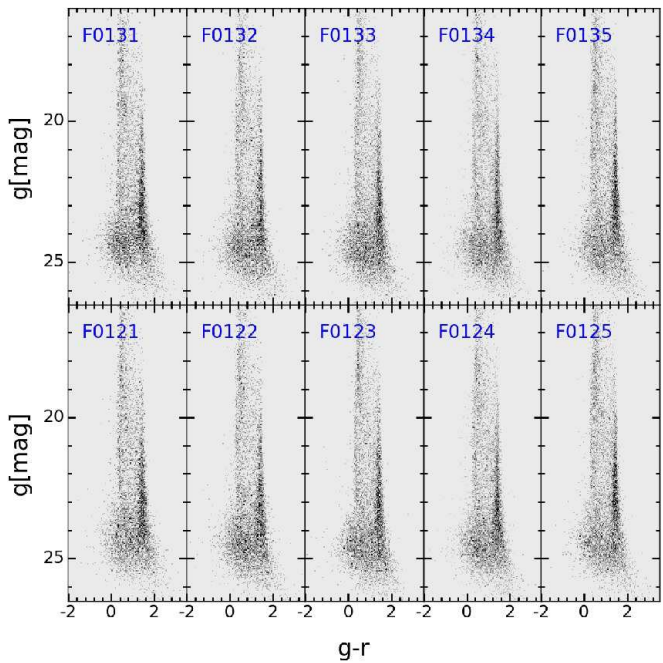}
\caption{$[g,g-r]$ colour--magnitude diagrams of 10 adjacent IAC Stripe
  82 fields covering 2.5 square degrees in total. Various sequences
  can be attributed to different Galactic populations like thin and
  thick disc or the stellar halo. The contamination by extended
  sources is small until $g$$\sim$23 mag. \label{fig.cmd}}
\end{figure}

\begin{figure}
\includegraphics[width=0.5\textwidth]{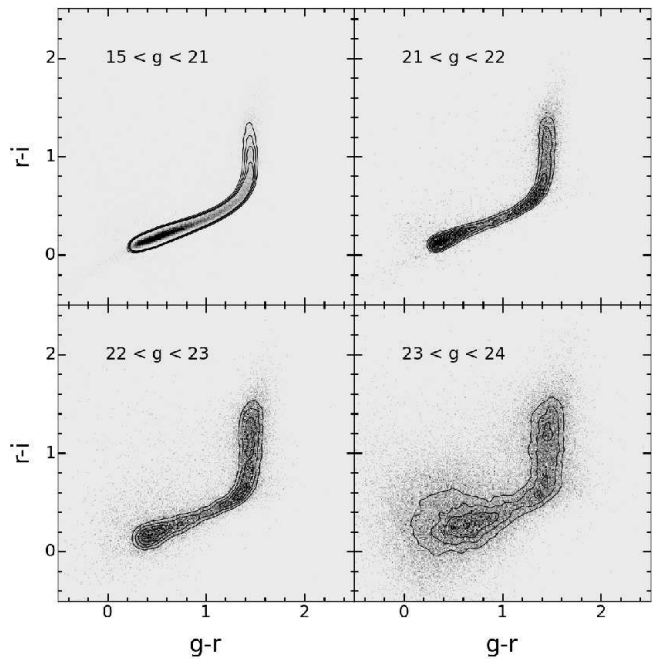}
\caption{$[g-r,r-i]$ colour--colour diagrams for point-like sources in 10
  square degrees of Stripe 82, split in different bins in $g$-band
  magnitude. For the contours a Gaussian smoothing of three times the
  binning size in colour (0.01~mag) has been applied. The contour
  levels mark two, three, four, five sources per binning interval. The stellar
  locus is well defined with only a small contamination by QSOs and
  extended sources until $g$$\sim$23~mag.\label{fig.color_color}}
\end{figure}

\subsection{Catalogues}
\label{sec.dr_cat}

We created catalogues of point and extended sources for the whole survey
area which are included in the release of the Stripe 82 data
products. The catalogues are based on {\tt SExtractor} photometry, they
include the positions, matched aperture luminosities, effective radii
and structural parameters like the moments of the brightness
distribution.

The catalogues contain objects being detected at least at the
3$\sigma$ level in $g$, $r$ and $i$, requiring 3 connected pixels
showing a 1.74$\sigma$ excess above the background in all three bands.
Objects are detected in each band separately; detections within a
search radius of 1 arcsec are merged to a common detection later
on. Kron apertures and structural parameters of the common detections
in $g$, $r$, and $i$ are determined in the deep $r$-band image which
yields a robust measurement of these parameters. To obtain the object
fluxes in matched apertures, {\tt SExtractor} is rerun in double-image mode
for each band on the Kron apertures measured in $r_{deep}$. We also
provide the object fluxes in fixed apertures of 5, 10, 20 and 30
pixels in diameter (1.98,3.96,7.92,11.88 arcsec),
useful to estimate the aperture correction for point sources. The
correction between the Kron magnitudes and the magnitude measured in
an infinite aperture depends on the waveband and the field within
Stripe 82, but usually is of the order of a few hundredth of a
magnitude between 0.02 and 0.05 mag.

\begin{figure}
\includegraphics[width=0.5\textwidth]{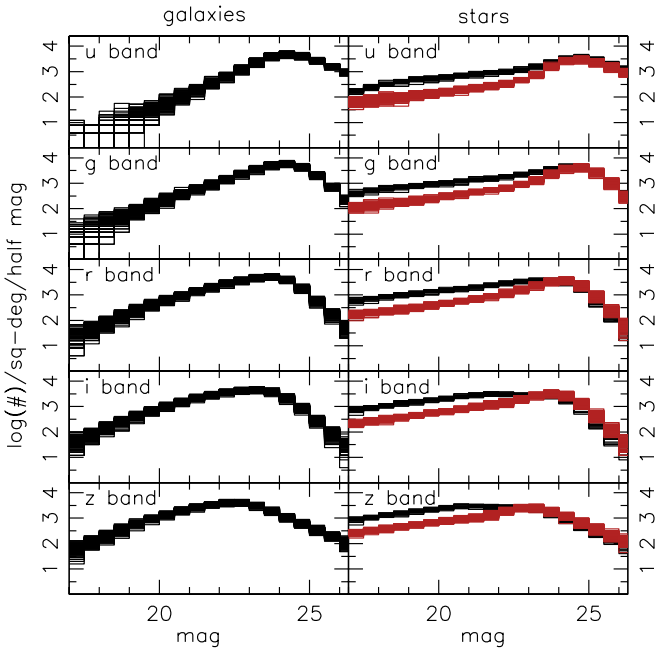}
\caption{Number counts of point and extended sources for 100 fields
  covering 25 square degrees in total. The fields are grouped around
  two different Galactic coordinates, ($l$,$b$)=(50\degr,--30\degr)
  (black lines, right-hand panel) and ($l$,$b$)=(70\degr,-50\degr) (red
  lines, right-hand panel ).
  \label{fig.numbercounts}}
\end{figure}

To separate point from extended sources, we rely on a
combination of {\tt SExtractor} runs in ($g,r,i,z,u$) and {\tt DAOPHOT~II/ALLSTAR}
\citep{Stetson1987} runs on the $r_{deep}$ images. Although of
slightly broader PSF than the $i$ or $z$ bands, the $r_{deep}$ images
are best suited for this task as the gain in depth allows to enhance
the magnitude limit for a proper separation of stars and
galaxies. Whereas {\tt SExtractor} has its advantages in the photometry of
extended sources, the {\tt DAOPHOT} package is optimized for point sources.
The combination of both methods allows, especially at fainter
magnitudes, a cleaner separation between extended and point-like
objects than one based on the {\tt CLASS\_STAR} or {\tt FLUX\_RADIUS} parameters
returned by {\tt SExtractor}. The {\tt DAOPHOT} runs yield an estimate of
stellarity or point-likeness encoded in the two parameters which are
output for each source, {\tt SHARP} (providing an estimate of the
concentration of the source) and {\tt CHI} (providing an estimate of
the goodness of fit of the object's profile to the stellar PSF). For
point-like sources we require $|${\tt SHARP}$|<0.5$ and
{\tt CHI}$<7.5$. Objects which fulfil these criteria in $r_{deep}$ are
assigned to the catalogue of point-like sources. All other objects
from the combined $g$, $r$, and $i$ catalogue are assigned to the
extended object catalogue.

Fig.~\ref{fig.cmd} shows colour--magnitude diagrams (CMDs) of
point-like sources for 10 adjacent fields. The CMDs show the
signatures of different Galactic populations like stellar halo and
thick disc at $g-r \sim 0.4$ or M stars in the thin disc, thick disc
or the halo at $g-r \sim 1.5$. The separation between stars and
galaxies works well until $g \sim 23$~mag. At fainter magnitudes the
contamination by extended objects is starting to get larger. To assess
the quality and the limit of the star/galaxy separation more closely,
we created colour--colour plots of the stellar catalogues. In
Fig.~\ref{fig.color_color}, we show $(g-r)$ plotted against $(r-i)$
for the point-like sources, split into different bins in $g$-band
magnitude. The stellar sequences get broader for fainter magnitude
bins due to larger photometric errors. They are well defined until
$g \sim 23$~mag with only a small fraction of outliers, the majority
of them presumably being mis-classified galaxies. Beyond this limit, the
contamination of the point source catalogue gets worse.

\subsection{Number counts}
\label{sec.dr_nc}

\subsubsection{Point dources}

In Fig.~\ref{fig.numbercounts}, we show the number counts of point-
and extended sources for 100 fields covering 25 square degrees in
total.  The fields are grouped around two different Galactic
coordinates at ($l$,$b$)=(50\degr,--30\degr) (black lines, right-hand panel)
and ($l$,$b$)=(70\degr,--50\degr) (red lines, right-hand panel). In the
stellar counts, this is reflected as overall decline in density for
higher latitudes.  The stellar counts are well described by a power-law
rise towards fainter magnitudes until the turn-over when
incompleteness starts to set in.  The bump-like excess of counts seen
in all bands, with its maximum at e.g. $g$$\sim$25~mag, is due to
galaxies being mis-identified as stars.  The values of the turn-over
magnitudes in the five bands agree with the results of the completeness
simulations. In these simulations we used the same detection criteria.
However, we did not require detections in multiple bands nor
included colours for the simulated objects.

\begin{figure}
\includegraphics[width=0.45\textwidth]{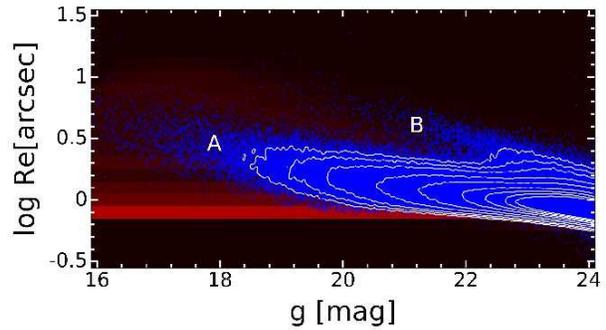}
\caption{Size--luminosity relation for extended objects in the
  $g$ band. Shown in blue are objects classified as non-point-like in
  the Stripe 82 catalogue. The labels `A' and `B' mark the main
  relation and the sequence of partially resolved galaxies, respectively.  
  The sources are overlaid on the completeness map for the simulation of
  exponential profiles, representing the recovery fraction based 
  on recovered sizes and luminosities of the input objects.
\label{fig.size_lum}}
\end{figure}

\subsubsection{Extended sources}

The number counts of galaxies are shown in the left-hand panel of
Fig.~\ref{fig.numbercounts}. The broadness of the number counts
distribution shows the field to field variations one can expect from
the Stripe 82 catalogues.  The presence of bright, saturated stars
and/or large galaxies affects the number counts in some co-adds. These
brighter objects affect the normalized counts in two ways, sometimes
by reducing the accessible area in the field, the other time by
enhancing the detection limit in their surroundings. Depending on the
brightness of the object, the affected areas could be several square
arcminutes in size. Due to smaller number densities at the bright end,
the variance of the galaxy counts is larger than the ones of the
stellar counts.

Another way to look at the galaxy populations is to plot their
effective radii against their luminosities. In Fig.~\ref{fig.size_lum},
we show the measured size--luminosity relation in the $g$ band. Most of
the galaxies fall on the main relation. A second, sparsely populated
sequence (labelled `B' in Fig.~\ref{fig.size_lum}) shifted towards
larger radii is populated by two classes of objects. The first class
consists of partially resolved nearby spiral galaxies which have been
split into multiple objects. The measured apertures trace spiral arms
or brighter knots like star-forming regions in these galaxies rather
than the galaxy as a whole. The second class populating this sequence
is made of faint galaxies in the haloes of brighter stars which have
both their size and luminosity been affected by the presence of the
bright object nearby. The main relation (labelled `A' in
Fig.~\ref{fig.size_lum}) follows the known correlation that brighter
galaxies tend to have larger effective radii. At small effective radii,
the relation saturates as the measured effective radii are limited by
the size of the PSF. This is also reflected in the completeness map
from our simulations, which is underlaid on the distribution of
measured galaxies. The completeness map gives the fraction of
recovered sources based not only on recovered position but also on
recovered sizes and luminosities. Comparing the completeness map with
the detection maps in Fig.~\ref{fig.completeness_ext} shows that
larger objects tend to be measured with smaller effective radii and
fainter magnitudes, a result of their lower surface brightness and 
the effect of the noise in the images. The prominent row at $\log(R_E)$=--0.1
in the completeness map is due to input sources with effective radii
smaller than the PSF size.

\section{Science showcases}
\label{sec.science}

\begin{figure*} 
\includegraphics[width=\textwidth]{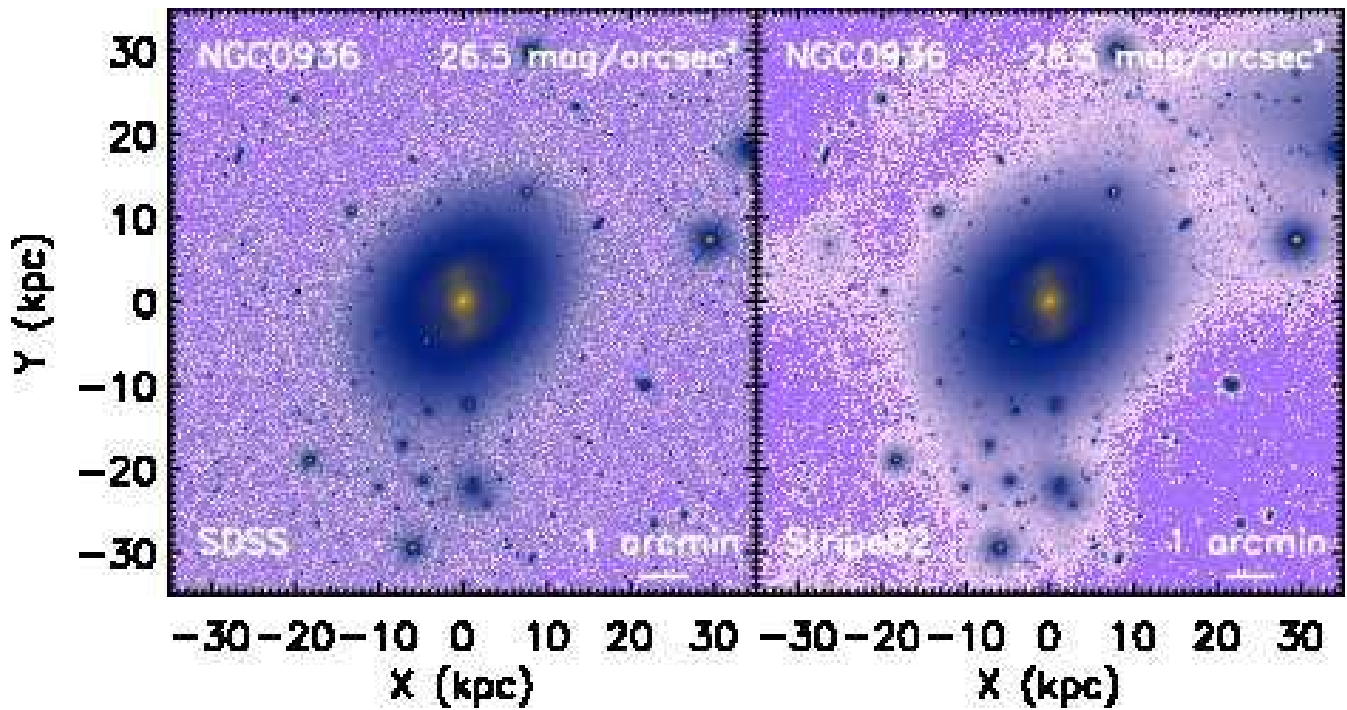} 
\caption{The effect of increasing the depth at observing nearby
  galaxies. The figure shows the galaxy NGC0936 (the so-called Darth
  Vader's galaxy) in the $r$ band.  The galaxy is shown at two
  different surface brightness limiting depths: $\mu_r$(3$\sigma$;
  10 arcsec $\times$ 10 arcsec) = 26.5 and 28.5 mag~arcsec$^{-2}$. The
  emergence of a faint tidal stream (27$\lesssim$$\mu_r$$\lesssim$28
  mag arcsec$^{-2}$) in the stellar halo of this galaxy exemplifies
  the need of ultra-deep observations to explore the substructure
  prediction by the $\Lambda$CDM model. \label{fig.ngc0936}}
\end{figure*}

As indicated in the introduction, a survey like Stripe 82 allows us to
address a number of astrophysical phenomena. In what follows, we will
illustrate some specific scientific cases that can be explored with
the reduction of the Stripe 82 we have conducted. These and other
questions will be fully analysed in forthcoming papers: stellar haloes
of disc and elliptical galaxies, tidal streams and satellite
populations, low surface brightness galaxies, ICL, 
optical cirrus, etc.

\subsection{Stellar haloes and disc truncations}
\label{sec.haloes}

Truncations in edge-on galaxies are known for a long time
\citep[e.g.][]{vanderKruit1979}, but have been proven elusive in
face-on galaxies. Two main families of models have been proposed for
the origin of truncations: suppression of star formation below a
certain gas density threshold for local instability
\citep{Fall1980,Kennicutt1989,Schaye2004}; and one in which the
truncation corresponds to a maximum in the specific angular momentum
distribution of the present disk, and might correspond to that in the
protogalaxy \citep{vanderKruit1987}.

The absence of stellar disc truncations in low-inclination spiral
galaxies has been a matter of debate in the last
decade. \citet{Martin-Navarro2014} argue that stellar haloes outshine
the galaxy disc at the expected position of the truncations, forcing
their studies to highly inclined (edge-on) orientations.  Using a
simple exponential disc plus stellar halo model based on current
observational constraints they show that truncations in face-on
projections occur at surface brightness levels comparable to the
brightness of stellar haloes at the same radial distance.

This view is supported by a recent study of halo truncations in 22
nearby, face-on to moderately inclined spiral galaxies, making use of
the $r_{deep}$ co-adds of our Stripe 82 reduction (Peters et al. 2016).
Truncations are found in only three galaxies, an
additional 15 galaxies are found to have an apparent extended stellar
halo.  Their simulations show that the scattering of light from the
inner galaxy by the PSF can produce faint structures resembling
stellar haloes, but in general this effect is insufficient to fully
explain the observed haloes.

There is a long debate about whether stellar haloes are formed in situ
during a monolithic collapse at early times \citep{Eggen1962} or are
built up by the remnants of accreted substructures
\citep{Searle1978}. Both models give very different predictions on
halo mean age and metallicities, age and metallicity gradients and the
overall density gradient. The deep Stripe 82 data reach depths which
allow, for the first time, to detect and characterize stellar haloes
for a statistically meaningful ($\sim$100) sample of galaxies of different types.
Studies like this will have immediate and important consequences for
our understanding of galaxy formation and evolution.

\subsection{Stellar streams}
\label{sec.streams}

\begin{figure*} 
\centering
\includegraphics[width=0.47\textwidth]{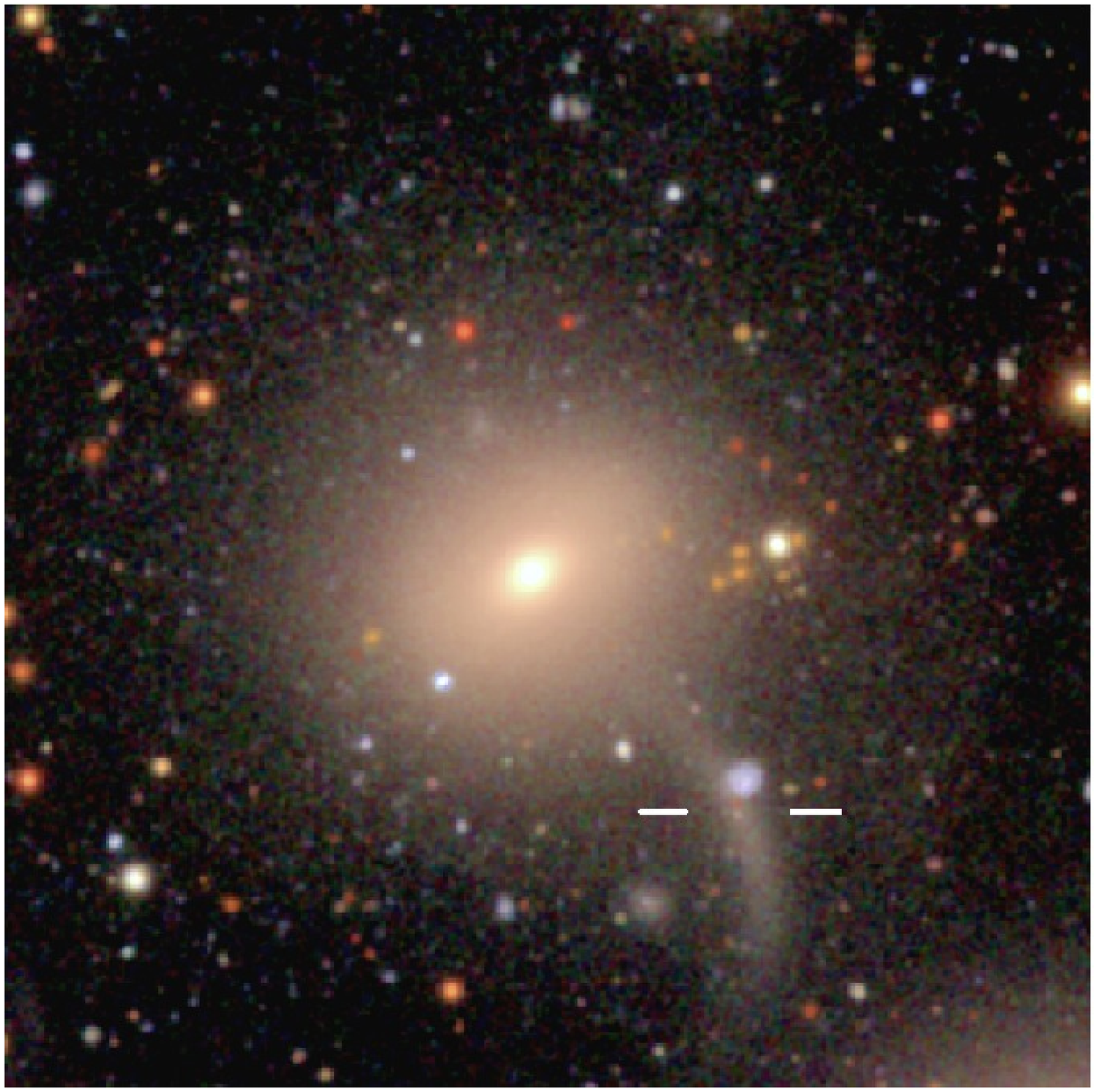}
\includegraphics[width=0.49\textwidth]{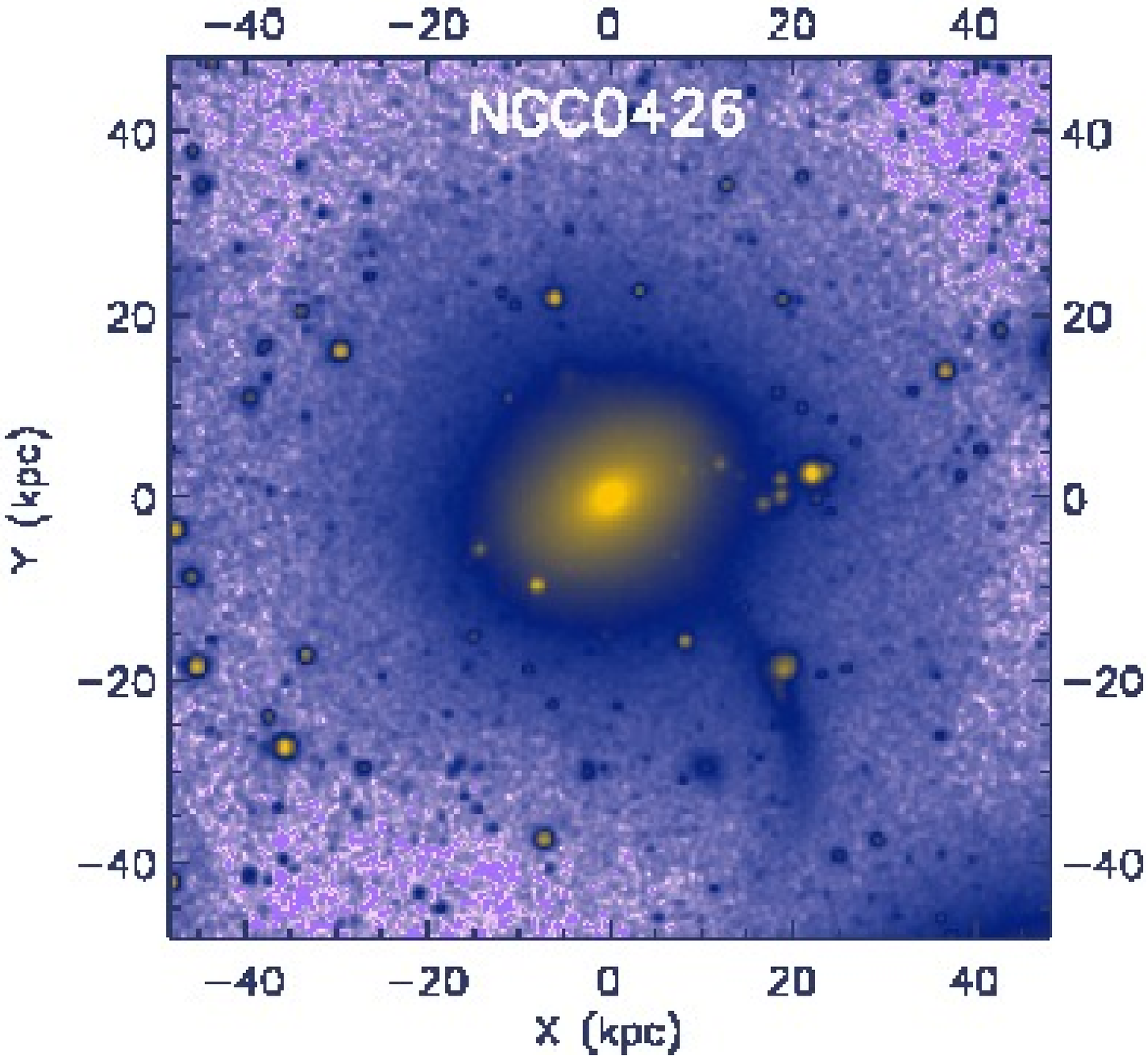}
\caption{Left-hand panel: real colour composite image of NGC0426 and
  its stream system \label{fig.ngc0426}. The main body of the dwarf
  satellite is still undergoing the disruption event. Two compact
  objects are located close to the suspected centre of the dwarf 
  (marked by white horizontal lines). The colours of
  the compact sources differ significantly, with one having a colour
  quite similar to the one of the stream. The second one is much redder,
  suggesting it being a background galaxy or a globular cluster.
  Right-hand panel: NGC0426 in the $i$ band, filtered with a median kernel of
  3 arcsec sidelength. The stream is fully embedded in the stellar
  halo and the shell system around the galaxy.}
\end{figure*}

$\Lambda$ cold dark matter ($\Lambda$CDM) models predict an increasing amount of substructure
(stellar streams, shells, filaments) within the stellar haloes of
galaxies when lowering the surface brightness threshold to values
below 30 mag arcsec$^{-2}$
\citep[e.g.][]{Bullock2005,Cooper2010,Font2011,Tissera2013}. Consequently,
the brightest streams will be in the reach of the surface brightness
limits of $\mu_r$ $\sim$28.5 mag arcsec$^{-2}$ we acquire in our
Stripe 82 co-adds. Here we show two exciting examples of the
possibilities and new discoveries Stripe 82 offers in this regard: an
extended faint stellar stream around NGC0936 (also known as the Darth
Vader's galaxy) and the ongoing disruption of a dwarf satellite by the
elliptical galaxy NGC0426.

The stream around NGC0936 (distance 19.8 Mpc) is seen as a huge loop
surrounding one third of the galaxy and extending $\sim$ 30 kpc from the
galaxy centre. With 27$\lesssim$$\mu_r$$\lesssim$28~mag arcsec$^{-2}$
and $(g-i)\sim 1.1$, the stream is close to the surface brightness
limit of our co-adds.  In Fig.~\ref{fig.ngc0936}, we show the stream,
comparing the depth of the single-pass SDSS DR7 imaging (left-hand panel)
versus the Stripe 82 $r$ band (right-hand panel), smoothed with a boxcar
average of 3~arcsec width. The surface brightness depth of the
images has been estimated using the rms of the images in boxes of
10$\times$10 arcsec$^2$ (i.e. around 10 $\times$ FWHM) and
corresponds to 3$\sigma$ detections. At its visible end, the loop
emerges into a diffuse object, presumably a dwarf galaxy undergoing
tidal disruption and the progenitor of the stream. The galaxy has a
bright core and seems quite extended, if the connection of the dwarf
and NGC0936 is confirmed its distance would imply a size of
$\sim$6~kpc in diameter. Its central surface brightness is
$\mu(r,0)$=21.4~mag arcsec$^{-2}$, the surface brightness just outside
the core is around $\mu_r$$\sim$26.3~mag arcsec$^{-2}$ with
$(g-i) \sim 1.0$. These properties match some of the characteristics
of the recently discovered class of ultra-diffuse galaxies (UDGs, see
Sec.~\ref{sec.lsb}). However, as this system is presumably tidally
disturbed, a relation to this type of objects is not straightforward.

As second example we show the stellar stream around NGC0426, a
redshift $z=0.017$ elliptical galaxy hosting an AGN (see
Fig.~\ref{fig.ngc0426}). The stream is relatively bright reaching
$\mu_g$ $\sim$24.9~mag arcsec$^{-2}$. In the Stripe 82 images, it is
traced from the centre of NGC0426 until projected distances of
$\sim$ 45 kpc. The system is seen in the state of ongoing disruption,
towards the centre of the presumed progenitor (at $x=18$~kpc, $y=-21$~kpc
in galactocentric coordinates) the images show two dense cores with
largely different colour characteristics. One of the compact cores has
a colour $(g-i) \sim 1.1$, only slightly redder than the colour of
the stream $(g-i) \sim 1.0$. The second core is much redder with
$(g-i) > 2$ which could point to a globular cluster or a background
galaxy seen projected on the stream.

\begin{figure}
\includegraphics[width=0.5\textwidth]{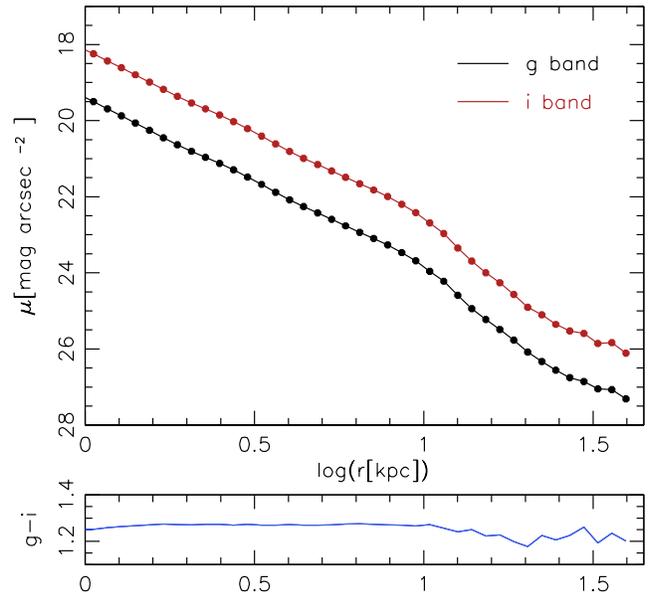} 
\caption{Surface brightness profiles of NGC0426 in the $g$ (black line)
  and $i$ (red line) bands. \label{fig.profile_ngc426}}
\end{figure}
We calculated surface brightness profiles for NGC0426 using the
{\tt IRAF/ELLIPSE} task. The profiles are shown in
Fig.~\ref{fig.profile_ngc426}, outside a radius of 2~arcsec (0.7
kpc) the colour is constant with $(g-i) \sim 1.3$. At an
isophotal radius $r$=31~arcsec (11 kpc), the profiles steepen;
beyond this radius, the ellipticity values decline from their
maximum at $\epsilon$=0.3 close to $\epsilon$=0 at $r$=60~arcsec
(21 kpc). The mean colour shows subtle changes after the break,
however remains redder than the main part of the stellar stream
throughout the profile. It is worth noting that a comprehensive
analysis of the colour gradient will require an exploration of the
effect of the PSF on the galaxy light distribution.  In
Fig.~\ref{fig.ngc0426} (right-hand panel), we show the $i$-band image of
NGC0426 which impressively reveals the extent of the stellar halo,
reaching to distances of $\sim$ 45 kpc and fully enclosing the stellar
stream and the disrupting dwarf galaxy. The halo is asymmetric with
a shell-like cut-off at its more extended right side. At least two
further objects could be connected to the NGC0426 system. The
galaxy located at ($x=19$~kpc, $y=-19$~kpc), slightly offset from the
presumed centre of the dwarf, seems to be too blue to act as
progenitor of parts of the material in the stellar halo. Also the
low surface brightness galaxy at ($x=10$~kpc, $y=-29$~kpc) shows a
bluer colour than the stellar halo ($(g-i) \sim 0.8$), although
the difference in colour is not too large.  At this stage we cannot
confirm nor rule out a connection of those objects with NGC0426 and
its stream system. Spectroscopic follow-up observations providing
information on distances and stellar populations will help to
disentangle the components and derive the recent accretion history
of NGC0426.
   
\subsection{Low surface brightness galaxies}
\label{sec.lsb}

Recently, a population of extremely diffuse galaxies with central
surface brightness 24$\leqslant$$\mu(g,0)$$\leqslant$26~mag
arcsec$^{-2}$ has been identified in the Coma cluster
\citep{Koda2015,vanDokkum2015}. One third of these roughly 1000
UDGs are Milky Way sized with large effective radii of $R_E > 1.5$~kpc,
however small stellar masses of
only few $10^7 - 10^8$~$M_{\sun}$. The UDGs have exponential light
profiles, effective radii $R_E \sim 0.8-5$~kpc and effective surface
brightness 25 $\leqslant$ $<$$\mu_e(R)$$>$ $\leqslant$ 28~mag
arcsec$^{-2}$.  Roughly, 8 per cent have compact nuclei at their centres. As
they show a distribution concentrated around the cluster centre it is
very likely that the great majority are indeed cluster members. This
is also supported by the fact that the objects are unresolved into
stars in agreement with them being at Coma's distance. The UDGs lie
along the red sequence in the CMD, consistent
with a passively evolving population, which may have lost their gas
supply at early times, possibly resulting in very high dark matter
fractions \citep{Yozin2015}.

The number density of UDGs in different environments is still unknown,
possibly most of these galaxies are confined to galaxy clusters where
several processes could help to remove the gas and quench star
formation. Indeed, first identifications of similar objects were
obtained in the Virgo cluster \citep{Sandage1984}. The galaxies which
were classified as `Huge Im or dE' are of huge size (up to 10 kpc in
diameter) and very low central surface brightness of about
$\mu(B,0)$=25~mag arcsec$^{-2}$. Also the 20 Virgo UDGs show a
concentration towards the centre suggesting a cluster membership. Only
a small number of large, low surface brightness galaxies are known in
the field \citep{Dalcanton1997,Burkholder2001,Impey2001}. Another
newly detected UDG recently reported in Mart\'inez-Delgado et al. (2016)
could reside in the field as well, although this object most probably
is  associated with a filament
of the Pisces-Perseus supercluster and not located in
isolation. The object lies at a redshift distance of 78 Mpc, implying
an extremely large size ($R_E \sim 4.7$~kpc) for its stellar mass of
only s few 10$^8$~$M_{\sun}$.  Its properties are consistent with
those found for UDGs; it shows a red colour ($V-I=1.0$), shallow
S\'ersic index ($n_V$=0.68), and no emission in H $\alpha$, typical of
dwarf spheroidal galaxies and suggesting that it is mainly composed of
old stars.

\begin{figure}
\includegraphics[width=0.5\textwidth]{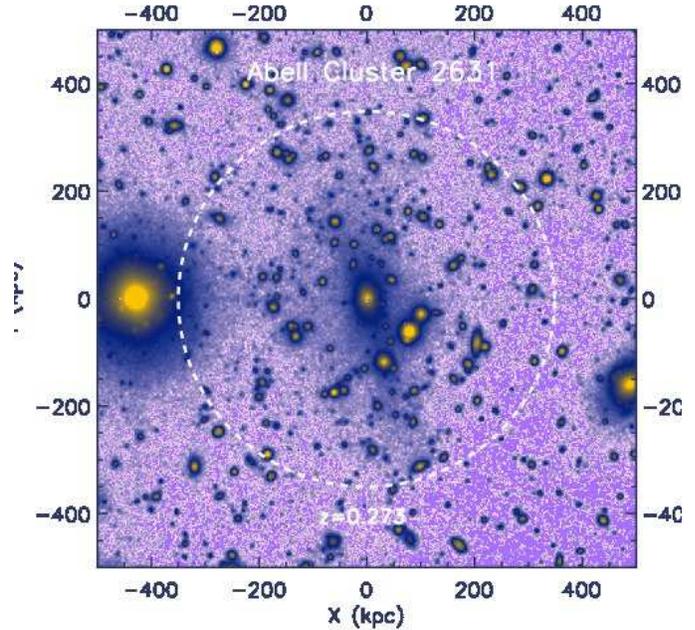}
\caption{Abell cluster 2631 at z=0.273 as seen in our Stripe 82
  $r$-band co-add.  The dashed line encircles a radius of 350 kpc. The
  image shows the extended ICL of the cluster observed
  down to the limit of the survey $\mu_r$$\sim$28.5~mag arcsec$^{-2}$.
  \label{fig.cluster}}
\end{figure}

Low surface brightness UDGs are well within the
detectability limit of the co-added Stripe 82 data, also proven by the
detection of the presumed progenitor galaxy of the stream around
NGC0936. This will allow one to probe their frequency in field
environments for a significant area, and will give new insights into their
formation and evolution. The large Stripe 82 area will further shed
some light on the abundance of predicted `dark galaxies' with surface
brightness even below the one of UDGs ($\mu_V$$>$27~mag
arcsec$^{-2}$).  These objects have been predicted within the
$\Lambda$CDM framework \citep[e.g.][]{Ricotti2005} but have never been
observationally sought, thus possibly remaining hidden from our
current surveys of the sky. Galaxies with such extreme low surface
brightness will be ideal tracers to explore the properties of almost
pure dark matter galaxies.

\subsection{Intra cluster light}

Little is known about the properties of the ICL
in galaxy clusters, the diffuse light component extending several
hundred kpc from the cluster centre with typical surface brightness
$\mu_V$ $\sim$26.5~mag arcsec$^{-2}$
\citep[e.g.][]{Mihos2005,Zibetti2005,Rudick2006}. The ICL is thought
to be made up primarily by stars which were tidally stripped from
their parent galaxy during interactions and mergers with other cluster
galaxies. The formation time-scales of the ICL are pretty much
unconstrained; measured metallicities show a wide spread from metal
poor to super-solar \citep[e.g.][]{Durrell2002,Krick2006,Montes2014}.
The Stripe 82 co-adds offer the possibility to address these questions
and characterize with an unparalleled quality the ICL properties: the
quantity of stars in this diffuse component, typical shapes and
extensions, and its relation to the entire mass of the galaxy
cluster. $\Lambda$CDM simulations predict around five massive clusters
($M_{Clu} > 10^{14}$~M$_{\sun}$) per square degree up to $z \sim$0.3
\citep{Hallman2007}. That means that within Stripe 82 it will be
possible to precisely derive the characteristics of the ICL in a
sample of around 1300 massive nearby galaxy clusters. Fig.~\ref{fig.cluster}
exemplifies the power of our Stripe 82 data for this
kind of studies. This figure shows the ICL in the
Abell cluster 2631 at $z=0.273$. The extension of the ICL
is clearly visible up to a radius of $\sim$350 kpc from the
cluster centre.

\subsection{Diffuse Galactic light}
\label{sec.cirrus}

\begin{figure}
\includegraphics[width=0.5\textwidth]{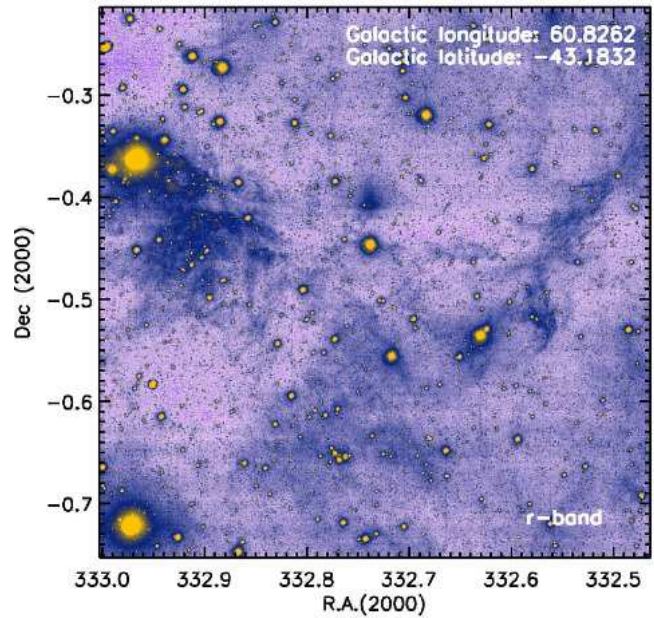}
\caption{A Stripe82 field of 32 arcmin $\times$ 32 arcmin size showing
  dust filamentary structure of our Galaxy in the $r$ band. The depth
  and resolution of our imaging allow us to explore faint optical
  cirrus with exquisite detail. \label{fig.cirrus}}
\end{figure}

First noticed in the late 1930s \citep[e.g.][]{Elvey1937,Henyey1941}, the
diffuse Galactic light or optical cirrus is starlight scattered off by
dust grains in the diffuse interstellar medium (ISM). Reprocessed
light is emitted at infrared wavelengths. The comparison between the
dust cirrus both in the optical and the infrared would give a powerful
tool for probing the dust properties as well as the interstellar
radiation field in the ISM
\citep[e.g.][]{Beichman1987,Guhathakurta1989}. The Stripe 82 area 
covers a large contiguous field of the sky ranging from Galactic
latitudes -65\degr$\lesssim$$b$$\lesssim$-25\degr. This allows one to
explore in large detail the optical variation of the dust properties
from regions closer to the Galactic plane to high Galactic
latitudes. Fig.~\ref{fig.cirrus} shows a
32 arcmin $\times$ 32 arcmin Stripe 82 field located at intermediate
Galactic latitude ($l$=60\degr.83, $b$=--43\degr.18). This $r$-band
image shows obvious dust filaments. The depth and resolution of the
Stripe82 are comparable with current optical studies focused on this
astrophysical phenomena \citep[e.g.][]{Ienaka2013} but the area
covered by the survey surpasses by a factor of $\sim$100 the explored
area in previous works. Undoubtedly, the analysis of the Stripe82 data
will become a major step forward in the understanding of the
ISM. Combining the optical emission with measurements in the UV
\citep[{\em GALEX;}][]{Niemack2009} and infrared
\citep[{\em Herschel;}][]{Viero2014} will establish the spectral energy
distribution of the cirrus for a large wavelength range. Its shape and
its dependences from both the emission strength and the intensity of
the radiation field will provide new insights into the emission and
heating mechanisms of the Galactic dust clouds. A closer insight into
the properties and characteristics of the Galactic dust emission is
urgently needed as its range in surface brightness intersects in large
parts with the brightness of the aforementioned science cases, from
stellar haloes and streams to the ICL. Indeed, Arp's
loop, a suspected stream-like ring around M81 suggesting an
interaction between M81 and M82, recently turned out to be likely a
superposition of a few recent star-forming regions located close to
M81, M81's extended disc and Galactic Cirrus emission
\citep{Davies2010,Sollima2010}, the latter being responsible for the
diffuse appearance of the feature.
 
\section{Conclusions}

The low surface brightness Universe only recently gathered more
attention and started to be investigated in greater detail. Stellar
streams around galaxies are now identified and characterized on a
regular basis using small- to medium-sized telescopes
\citep[e.g.][]{Martinez2008,Martinez2010,Duc2014,vanDokkum2014} .
Most of these surveys use pointed observations of pre-selected targets
and are restricted to a limited field of view.

The Stripe 82 data cover 275 square degrees and reach surface
brightness limits $\mu_r$$\sim$28.5~mag arcsec$^{-2}$ (3$\sigma$,
10 arcsec $\times$ 10 arcsec), comparable to the one's acquired in the
aforementioned studies. In this way, it can be used as a blind survey
allowing statistical tests of the frequency and characteristics of low
surface brightness phenomena like UDGs, stellar streams, extended
discs or tidal interactions for different types of galaxies.

In our reduction of the Stripe 82 data, we put special emphasis on
preserving the characteristics of the low surface brightness
structures on all spatial and intensity scales.  Our effort is made
publicly available through a webpage at
http://www.iac.es/proyecto/stripe82.  The data release includes the
co-added data in 5+1 bands $(u,g,r,i,z,rdeep$=$<$$g$+$r$+$i$$>$),
corresponding exposure time maps and image representations of the
PSF. We also provide separate source catalogues for point and extended
objects with stars and galaxies confidently separated until
$g  \sim 2$3~mag. Our completeness simulations showed that for
exponential profiles the data are 50 per cent complete at the $3\sigma$ level
for an effective surface brightness $<$$\mu_e(r)$$>$$\sim$25.5~mag
arcsec$^{-2}$. The 50\% completeness limits for point sources are
(24.2, 25.2, 24.7, 24.3, 23.0)~mag in $(u,g,r,i,z)$, reaching between 0.1
and 0.3 mag deeper than the reduction of \citet{Annis2014} while being
at the same depth as \citet{Jiang2014}.

Discoveries like the faint stellar streams around NGC0936 and NGC0426
in our data give a glimpse of the prospects of deep wide-field surveys
like Stripe 82 for the study of the, until now almost hidden, low
surface brightness Universe. The full exploration of the Stripe 82
data set, including its extensions to wavebands from the UV to the
infrared, will have the possibilities to add significant pieces of new
insights to this relatively new research field.

\section*{Acknowledgements}
We thank the anonymous referee for helpful comments. We further thank
Lee Kelvin, Mauricio Cisternas, Mar\'ia Cebri\'an, Mireia Montes and
Javier Rom\'an for tests of the data base.  This work has been supported
by the Programa Nacional de Astronom\'ia y Astrof\'isica of the
Spanish Ministry of Science and Innovation under grant MINECO
AYA2013-48226-03-1-P.

Funding for SDSS-III has been provided by the Alfred P. Sloan
Foundation, the Participating Institutions, the National Science
Foundation and the US Department of Energy Office of Science. The
SDSS-III web site is http://www.sdss3.org/. SDSS-III is managed by the
Astrophysical Research Consortium for the Participating Institutions
of the SDSS-III Collaboration including the University of Arizona, the
Brazilian Participation Group, Brookhaven National Laboratory,
University of Cambridge, Carnegie Mellon University, University of
Florida, the French Participation Group, the German Participation
Group, Harvard University, the Instituto de Astrof\'{\i}sica de
Canarias, the Michigan State/Notre Dame/JINA Participation Group,
Johns Hopkins University, Lawrence Berkeley National Laboratory, Max
Planck Institute for Astrophysics, Max Planck Institute for
Extraterrestrial Physics, New Mexico State University, New York
University, Ohio State University, Pennsylvania State University,
University of Portsmouth, Princeton University, the Spanish
Participation Group, University of Tokyo, University of Utah,
Vanderbilt University, University of Virginia, University of
Washington and Yale University.

\bibliography{s82_data}{}
\bibliographystyle{}

\end{document}